\documentclass{pasa}%

\usepackage{graphicx}
\usepackage{rotating}

\title{Megahertz emission of massive early-type stars in the Cygnus region}

\author[Benaglia et al.]{P. Benaglia$^{1,2}$, M. De Becker$^3$, C. H. Ishwara-Chandra$^4$, H. Intema$^{5,6}$ and N. L. Isequilla$^2$
\affil{$^1$Instituto Argentino de Radioastronomia, CONICET \& CICPBA, CC5 (1897) Villa Elisa, Prov. de Buenos Aires, Argentina}%
 \affil{$^2$Facultad de Ciencias Astron\'{o}micas y Geof\'{\i}sicas, UNLP, Paseo del Bosque s/n, (1900) La Plata, Argentina}
 \affil{$^3$Space sciences, Technologies and Astrophysics Research (STAR) Institute, University of Li\`ege, Quartier Agora, 19c, All\'ee du 6 Ao\^ut, B5c, B-4000 Sart Tilman, Belgium}
 \affil{$^4$National Centre for Radio Astrophysics (NCRA-TIFR), Pune, 411 007, India}
 \affil{$^5$International Centre for Radio Astronomy Research, Curtin University, Bentley, WA 6102, Australia}
 \affil{$^6$Leiden Observatory, Leiden University, Niels Bohrweg 2, 2333 CA Leiden, the Netherlands}
}%

\jid{PASA}
\doi{10.1017/pas.\the\year.xxx}
\jyear{\the\year}

\usepackage{aas_macros}
\usepackage{hyperref} 
\hypersetup{colorlinks,citecolor=blue,linkcolor=blue,urlcolor=blue}

\hypersetup{draft}

\begin{document}

\begin{frontmatter}
\maketitle

\begin{abstract}
{Massive, early type stars have been detected as radio sources for many decades. Their thermal winds radiate free-free continuum and in binary systems hosting a colliding-wind region, non-thermal emission has also been detected. To date, the most abundant data have been collected from  frequencies  higher than  1~GHz. We present here the results obtained from observations at 325 and 610~MHz, carried out with the Giant Metrewave Radio Telescope, of all known Wolf-Rayet and O-type stars encompassed in area of $\sim$15~sq degrees centred on the Cygnus
region. We report on the detection of 11 massive stars, including both Wolf-Rayet and O-type systems. 
The measured flux densities at decimeter wavelengths allowed us to study the radio spectrum of the binary systems and to propose a consistent interpretation in terms of physical processes affecting the wide-band radio emission from these objects. WR\,140 was detected at 610~MHz, but not at 325~MHz, very likely because of the strong impact of free-free absorption. We also report -- for the first time -- on the detection of a colliding-wind binary system down to 150~MHz, pertaining to the system of WR\,146, making use of complementary information extracted from the TIFR GMRT Sky Survey. Its spectral energy distribution clearly shows the turnover at a frequency of about 600~MHz, that we interpret to be due to free-free absorption. Finally, we report on the identification of two additional particle-accelerating colliding-wind binaries, namely Cyg\,OB2\,12 and ALS\,15108\,AB.}
\end{abstract}

\end{frontmatter}

\section{INTRODUCTION}
\label{sec:intro}
Massive, early-type stars -- OB and Wolf-Rayet (WR) classes -- are characterized by strong winds that deliver material and energy to their surroundings. Many of these stars are in binary or higher multiplicity systems. In such systems, stellar winds interact in a colliding-wind region (CWR), hence their designation as colliding-wind binaries or CWB. Since the $'$70s massive stars are known to be moderate sources of thermal radio emission, and in many cases non-thermal (NT) radio emission has also been identified. The first is naturally explained as free-free radiation from the stellar winds \citep{WB}. The NT contribution arises from synchrotron radiation produced by relativistic electrons in the presence of the local magnetic field in the CWR  \citep[e.g.][]{doughwilliams2000}. Consequently, NT radio emission constitutes a valuable tracer of particle acceleration in these systems. The mechanism responsible for the acceleration of relativistic electrons is very likely diffusive shock acceleration \citep{1983RPPh...46..973D,EU}. CWBs { emit} radio waves but are dim and scarce. Only the closer { systems} can be detected with current facilities, and intensity distribution maps could be obtained for a handful of them only \citep[][and references therein]{Benaglia2015,SanchezBermudez2019}. As the NT emission arises from the CWR, a significant variability on the orbital time scale is expected, and confirmed by adequate radio monitoring, such as in the case of the emblematic system WR\,140 \citep{WhBe140}. The most complete catalogue of particle-accelerating colliding-wind binaries (PACWBs) was published recently by \citet{catapacwb}, and contains around 40 objects and a complete compilation of properties of the objects along the entire electromagnetic spectrum.

On the one hand, only a limited fraction of massive early-type stars or systems have been observed in the 1--15~GHz frequency range, most of which with flux densities of the order of a few mJy. The brightest O-type PACWB is HD~167971 with a few tens of mJy \citep{Blomme2007}, and the absolute record { for a  PACWB} so far is held by the recently discovered Apep system (WR+WR binary), with flux densities higher than 100~mJy \citep{callingham2019}. On the other hand, observations at very low frequencies have been, up to know, very limited, due to instrumental constraints, mainly angular resolution but also sensitivity. The main antecedent to this work is that of \citet{Setia2003}. The authors scrutinized a $2^\circ \times 2^\circ$ area of Cygnus\,OB2 at 1.4~GHz and 350~MHz using the Westerbork Synthesis Radio Telescope { (WSRT)}, to investigate hot massive stars. Their 350-MHz observations reached an angular resolution of about $55''$, and an average 1-$\sigma$ { flux density} value of 2 -- 3~mJy and resulted in the detection of three  Wolf-Rayet systems. 

More recently, the Giant Metrewave Radio Telescope (GMRT) has proven its efficiency to study such stellar systems: \citet{Benaglia2019} reported on the study of WR\,11 by means of GMRT data from 150~MHz to 1.4~GHz, with detections at 325 and 610~MHz at angular resolutions ranging from 22 to 5$''$. The synchrotron emission is stronger at lower frequencies, hence the relevance to consider GMRT observations to investigate the non-thermal physics of these systems \citep{DeBecker2019}. At frequencies lower than 1~GHz, turnover processes such as Razin-Tsytovich effect (RTe), free-free (thermal) absorption (FFA) and synchrotron self-absorption (SSA) may shape the spectrum depending on various factors that differ from one system to another. 
The characterization of this turnover, i.e. its nature and its occurrence frequency, constitutes to date a poorly investigated part in the description of the radio behaviour of massive star systems. It is clear that FFA is likely to be the dominant turnover process, because of the presence of the optically thick wind material. Characterizing this FFA is important for a confrontation to hydro-radiative models of CWBs (including system orientation and geometry). However, if the spectrum inversion is due to RTe, the turnover frequency is related to the number density of thermal electrons and to the magnetic field strength in the emission region. SSA is less likely as it would require number densities in relativistic electrons certainly higher than expected in CWBs.

At the other end of the electromagnetic spectrum, significant progress has been made to model the emission from colliding-wind regions in massive binaries \citep[see][and references therein]{delPalacio2016}. The high-energy non-thermal photon emission { has recently been} modeled in three dimensions, taking into account particle acceleration and radiative losses leading to $\gamma$-ray emission \citep{Reitberger2014a,Reitberger2014b,Reitberger2017}. However, high-energy emission from PACWBs lies mainly below the detection thresholds of the current instrumentation. The only exceptions are $\eta$\,Car \citep{Tavani2009,Abdo2010,Reitberger2015} and WR\,11 \citep{Pshirkov2016,WR11variab}. For this reason, radio observations constitute the most adequate approach to investigate these systems, especially from the point of view of their non-thermal physics. 

We have carried out observations with the GMRT at two decimeter bands, on a region of $\sim$15~sq~deg, centred at $RA,Dec$(J2000) = 20:25:30, 42:00:00, that belongs to the Cygnus Rift, in the northern sky. We reached an angular resolution down to 10$''$ and average rms down to 0.5 mJy beam$^{-1}$. These last numbers mean roughly an improvement in factor 5 both in angular resolution and sensitivity with respect to the previous survey of Cyg\,OB2 carried out with the WSRT.
Some parts of this region were previously observed with the GMRT to investigate potential counterparts to TeV sources \citep{Marti2007,Paredes2007}, but this is the first time such a wide area of the Cygnus region is studied at sub-GHz frequencies with the mentioned settings. 
The region under study here is populated by plenty of WR and OB stars, including the archetypal systems WR\,140, WR\,146, WR\,147 and Cyg\,X-3, and our project allowed us to measure the radio emission at their stellar positions. In Section\,\ref{cygnus}, we briefly describe the observed region and its massive stellar content. Section\,\ref{obsdata} presents the observing details and data reduction. In Section\,\ref{results}, we show the results of the performed analysis. Our results are then discussed in Section\,\ref{disc}, and we finally conclude in Section\,\ref{concl}. 

\begin{figure}[t]
\begin{center}
\includegraphics[width=8cm,angle=-0]{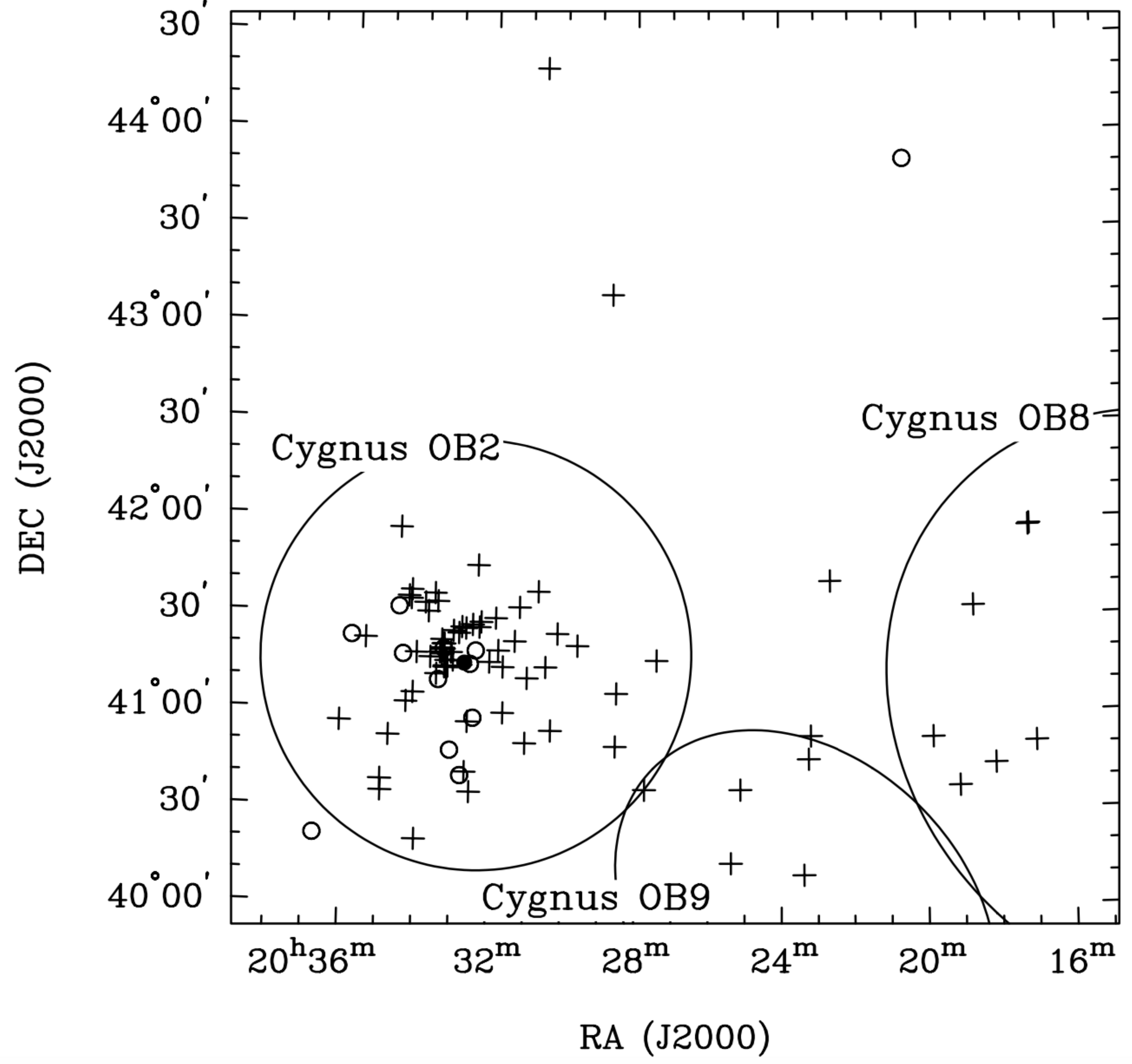}
\caption{Observed area, showing the positions of the stars studied here, with open circles (detected cases) and crosses (undetected cases). The large ellipses mark the  approximate extensions of the OB associations present in the area, according to \citet{uyaniker2001}.}
\label{Figfield}
\end{center}
\end{figure}

\begin{sidewaystable}
\caption{Wolf-Rayet stars in the observed field: measured radio { flux density} or upper limit, and spectral index information.}
\label{fluxWR}
\centering
\begin{tabular}{l@{~~~}l@{~~~}c@{~~~}r@{~~~}r@{~~~}r@{~~~}r@{~~~}r@{~~~}r}
\hline
Name &  Spectral type & $RA,Dec_{\rm J2000}$ & $d$  & $S_{\rm 610MHz}$ & Observing & $S_{\rm 325MHz}$ & Observing & $\alpha_{325}^{610}$ \\
     &  classification &  (h,m,s),(d,','') & (kpc) & (mJy)  & date(s) & (mJy) & date(s) & \\
\hline
WR\,138a & WN8-9h$^{(1)}$ & 20:17:08.12, $+$41:07:27.0 & 12.5$^\dag$ & $<$0.5 & 18/06/2015 & $<$1.5 & 7/02,26/09/2015 &---\\
WR\,140  & WC7pd$+$O5.5fc$^{(2)}$ &20:20:27.98, $+$43:51:16.3 & 1.64$^\dag$ & 0.87$\pm$0.10 & 17/07/2016 & $<$0.5 & 6/02/2015 & $>$\,0.8\\ 
WR\,142a &WC8$^{(3)}$ & 20:24:06.19, $+$41:25:33.8  & 1.81$^\dag$ & $<$1.0 & 17/08/2015 &$<$1.75 & 26/10/2014 & ---\\
WR\,142b &WN6$^{(4)}$ & 20:28:14.56, $+$43:39:25.5 & 1.77$^\dag$ &  $<$0.75 & 1/07/2016 & $<$1.25 & 6/02/2015 & ---\\
WR\,144  &WC4$^{(5)}$ & 20:32:03.02, $+$41:15:20.5 & 1.75$^\dag$ & $<$0.75 & 14/07,8/08/2016 & $<$1.25 & 4/11/2013 &---\\
WR\,145  &WN7o/CE$+$O7V((f))$^{(6)}$ & 20:32:06.28, $+$40:48:29.5 & 1.46$^\dag$ & $<$0.75 & 29/11/2014 & $<$1.5 & 4/11/2013 &---\\
WR\,145a &WN4-6$+$CO$^{(7)}$ & 20:32:25.78, $+$40:57:27.9 & 7.4\,\,\, & 66 $\pm$0.5 & 29/11/2014,14/07,8/08/2016& 27$\pm$0.5& 4/11/2013 & 1.42$\pm$0.07 \\
WR\,146  &WC6e$+$O8I-IIf$^{(8)}$ & 20:35:47.09, $+$41:22:44.7 & 1.1$^\dag$ & 121$\pm$4 & 29/11/2014& 111$\pm$2 & 4/11/2013 & 0.1$\pm$0.04\\ 
WR\,147  &WN8(h)$+$B0.5V$^{(9)}$ &20:36:43.60, $+$40:21:08.0 & 1.79$^\dag$ & 20.6$\pm$0.3 & 11,21/08/2016& $<$20 & 4/11/2013 & $>$\,0 \\ 
\hline\hline
\end{tabular}
\tabnote{References: (1) \cite{Gvaramadze2009}; (2) \cite{Fahed2011}; (3) \cite{Pasquali2002}; (4) \cite{Littlefield2012}; (5) \cite{Sander2012}; (6) \cite{Muntean2009}; (7) \cite{KM2017}; (8) \cite{Lepine2001}; (9) \cite{williams1997}. $\dag$: \cite{rate2020}. $RA$, $Dec$ (J2000): optical positions.}
\end{sidewaystable}

\begin{sidewaystable}
\caption{Measured radio { flux density} or upper limit, and spectral index information of the detected O-type stars in the observed field.}
\label{fluxO}
\centering
\begin{tabular}{l@{~~~}l@{~~~}c@{~~~}c@{~~~}r@{~~~}r@{~~~}r@{~~~}r@{~~~}r}
\hline
Name &  Spectral type & $RA_{\rm J2000}$   & $Dec_{\rm J2000}$    & $S_{\rm 610MHz}$ & Observing & $S_{\rm 325MHz}$ & Observing & $\alpha_{325}^{610}$ \\
     &  classification &  (h,m,s)  & (d,','') & (mJy)  & date(s) & (mJy) & date(s) & \\
\hline
Cyg\,OB2-5  & (O6.5:Iafe $+$ O7Iafe) $+$OB & 20 32 22.42&  +41 18 19.0 & 4.0$\pm$0.15 & 14/07,8/08/2016 & 5.5$\pm$0.3 & 4/11/2013  & $-$0.52$\pm$0.27 \\
  & + O7 Ib(f)p var?$^{(1)(2)}$ &  &  &  &  &  \\
Cyg  OB2-A11  & O7Ib(f)$^{(3)}$ & 20 32 31.54 & +41 14 08.21 & 0.5$\pm$0.1 & 14/07,8/08/2016 & $<$1.5 & 4/11/2013 & $>$$-$1.9\\  
ALS\,19624   & O8II((f))$^{(4)}$ & 20 33 02.92 & +40 47 25.3 &  0.5$\pm$0.1 & 29/11/2014 & $<$1.5 & 4/11/2013 & $>$$-$1.6\\
Cyg\,OB2-8A & O6Ib(fc)\,$+$\,O4.5:III:(fc)$^{(3)}$ & 20 33 15.08 &+41 18 50.51 & 2.4$\pm$0.2 &  14/07,8/08/2016 & $<$2.3 & 4/11/2013 & $>$\,0.1\\
ALS\,15108\,AB  & O6IV((f))$^{(3)}+$companion? & 20 33 23.46 & +41 09 13.0 & 0.95$\pm$0.2 & 14/07,8/08/2016 & 1.25$\pm$0.2 & 4/11/2013 & $-$0.44$\pm$0.96\\
Cyg OB2-73 &  O8Vz\,$+$\,O8Vz$^{(3)}$ & 20 34 21.93 & +41 17 01.7 & 0.4$\pm$0.1 & 14/07,8/08/2016 & $<$2 &4/11/2013 & $>$$-$2.5\\ 
Cyg\,OB2-335 & O7V((f))\,$+$\,O7IV((f))$^{(5)}$ &20 34 29.60 &+41 31 45.5 & 5.4$\pm$0.2 & 29/11/2014 & 2.6$\pm$0.3 & 4/11/2013 & 1.16$\pm$0.44 \\
\hline\hline
\end{tabular}
\tabnote{References for spectral-type classification: (1) \cite{MONOS}; (2) \cite{Kennedy2010}; (3): \cite{GOSS2016}; (4) \cite{comeron2012}; (5) GOSC v4.2 (https://gosc.cab.inta-csic.es/). $RA$, $Dec$ (J2000): optical positions.}
\end{sidewaystable}
 
\section{THE REGION AND SOURCES UNDER STUDY}\label{cygnus}

The sky area known as the Cygnus constellation ($-65^\circ \leq l \leq 95^\circ, -8^\circ \leq b \leq +8^\circ$) contains active star-forming regions relatively close-by (d $\leq$ 2.5 kpc). Despite that the high absorption in its line of sight prevents to accurately map the stellar population at optical wavelengths, Cygnus stands as one of the richest and more crowded areas in the Galaxy regarding massive stellar objects \citep[see the review paper by][]{reipurthcyg}. It houses nine OB associations together with bright open clusters \citep{uyaniker2001,Mahy2013}. 
Among the stellar associations, Cyg\,OB2 is one of the youngest groups, with one hundred O stars and thousands of B stars \citep[e.g.][]{knodl}. At its side, Cyg\,OB1, Cyg\,OB8 and Cyg\,OB9 harbor about a hundred hot stars. We have carried out an investigation of the low frequency emission at the position of the massive early-type stars belonging to the mentioned associations. Figure\,\ref{Figfield} presents the region surveyed here.

A first search for WR and O-type stars along the region portrayed in Fig.\,\ref{Figfield}, using the Simbad database, resulted in nine WR and around 90 O-type stars. The search was refined consulting the Wolf-Rayet list 
at http://www.pacrowther.staff.shef.ac.uk/WRcat/ \citep[see also][]{rate2020} and the last version [GOSC v4.2] of the Galactic O-star catalog (https://gosc.cab.inta-csic.es/) \citep[see][along with its references]{MONOS}. The WR stars list is given in Table\,\ref{fluxWR}. The O-type stars considered here are listed in Table\,\ref{fluxO}, and Table\,\ref{nondet} of Appendix A. 
Out of 87 O-type stars and systems, 50 of them are listed as GOSC members, { while 28 appear in \cite{comeron2012}.}

\section{GMRT OBSERVATIONS AND DATA REDUCTION}\label{obsdata}

The observations were collected along four campaigns (2013, 2014--2015, 2015 and 2016--2017) using the Giant Metrewave Radio Telescope, at two bands -centred at 325 and 610~MHz-, and cover a total sky area of about { 15} square degrees.
We used total intensity mode, with a bandwidth of 32~MHz and consisting of 256 spectral channels to minimise bandwidth smearing. As the common practice, we observed the flux { density} calibrators (3C286, 3C48 and 3C47) at the beginning and end of each  run, and a phase calibrator 2052$+$365 for five minutes, bracketing the 30-min scans of the targeted FoVs.
Five pointings were performed in the 325~MHz band and forty-seven in the 610~MHz band.
The total observing time was 172 hours which were divided so as to achieve near uniform noise in the survey. The fields of view of GRMT are $81'\pm4'$ and $43'\pm3'$ at 325~MHz and 610~MHz, respectively\footnote{GMRT Observer’s Manual; www.ncra.tifr.res.in/ncra/gmrt /gmrt-users/observing-help/manual\_7jul15.pdf}. Additional details in the observing process, like FoV centres, exact dates of observation and name of calibrators used are given in Benaglia et al. (2020, submitted).

The data were reduced by means of the SPAM routines \citep{IntemaSPAM}. The SPAM pipeline allowed calibration and flagging of each FoV. It also accounted for the $T_{\rm sys}$ correction for excess background towards the science target. 
For the 325-MHz FoVs the correction factor ranged from 1.7 to 3.6, and for 610~MHz, from 1.22 to 1.75, the larger values near the galactic plane. The process included several rounds of self-calibration to the antenna phases. The images were primary beam corrected and finally combined in a mosaic, weighting them with the inverse of the variance.

The images were built with robust weighting of $-$1, since we were more interested in discrete sources.  Thus extended emission could not be imaged at its best. The used settings combined with the complex diffuse structure in the Cygnus region yielded to variations in the rms across the images. The average rms attained in the final mosaics is 0.5~mJy per beam at 325~MHz, and 0.2~mJy per beam at 610~MHZ. The synthesized beams were set to $10'' \times 10''$ and $6'' \times 6''$.

\section{RESULTS}\label{results}

We looked for radio emission at the location of the WR and O-type stars in the observed region.
We claimed a detection if the optical position of the star/system is located within the radio emission loci (i.e. the size of the synthesized beam for discrete radio sources), and if the radio { flux density} remained above three times the rms of the surrounding area. We list the detection  results on  the  WR stars in Table\,\ref{fluxWR}, including upper limits (3rms) of the flux density for the undetected cases. The { flux densities} related to the detected O-type stars are given in Table\,\ref{fluxO}. The detected radio sources at the positions of WR and O-type stars are presented in Figures 2 to 12.

To measure the flux density we either fitted a Gaussian function with a background linear level if needed, or considered the { flux density} peak for discrete sources. The rms for each region was estimated by averaging the noise over four boxes surrounding the stellar position, free of sources.

Additionally, we searched for 150-MHz radio emission at the positions of the stars from Tables\,\ref{fluxWR} and \ref{fluxO} using the TIFR GMRT Sky Survey Alternative Data Release~1 \citep[TGSS ADR1,][]{IntemaTGSSADR1}, and reprocessing of these data (H. Intema, private comm.). The rms levels kept in the range 8--10~mJy~beam$^{-1}$, except at the position of WR\,138a that doubled that value. 

One HMXB, WR\,145a, and three well known colliding-wind systems (WR\,140, WR\,146 and WR\,147) were detected. WR\,145a is composed by a WN star and a compact object, at a distance of 7.4~kpc, and with a period of 0.2 d \citep{Zdziarski2018}, and will not be further analyzed here since the nature of the radio emission is out of the scope of this work. The system of WR\,140 was detected at 610~MHz, at phase 0.94 of its 7.93-yr period \citep[see][and references therein for information about this well monitored system at other radio frequencies and electromagnetic ranges]{Dougherty2005}. The orbital phase corresponding to our observations was derived using the ephemeris of \citet{monnier2011}. WR\,146, with a period of many years, hosts a colliding-wind region; the latest estimate  of its distance is 1.1~kpc. \citet{Hales2016} preformed on it a prototype  full-polarimetric study that yielded no fractional polarization of its radio emission above 0.6\% (see also references therein). We 
detected it for the fist time at 150 and 610~MHz. 
The CWB WR\,147, according to \citet{rate2020}, is located 1.79~kpc from us, a  distance three times larger than the former largely accepted value. It follows a long orbit; its radio spectrum was studied by \citet{Skinner1999} and \cite{Setia2001}. We report its detection at 610~MHz. At 325~MHz, its location, at the border of the FoV, precluded detection below 20~mJy.

Among the O-type stars, Cyg\,OB2-5 is a system made of at least four components, with probably more than one colliding-wind region \cite[see][and its references]{dzib2013}. { Flux density} changes are expected along the orbital phases. \citet{Marti2007} already reported a detection of this system at 610\,MHz, with a flux density of 2.41\,$\pm$\,0.22~mJy, so significantly lower than our measurement of 4.0\,$\pm$\,0.15~mJy. Regarding the binary Cyg\,OB2-8A, \cite{Blomme2010} showed that the radio emission is phase locked with the orbital period of 22 d. The radio emission of Cyg\,OB2-335 was discovered by \citet{Setia2003}, who proposed it as a CWB, based on observations at 1.4, 4.9 and 8.4~GHz.

\begin{figure}[!h]
\begin{center}
\includegraphics[width=7cm,angle=0]{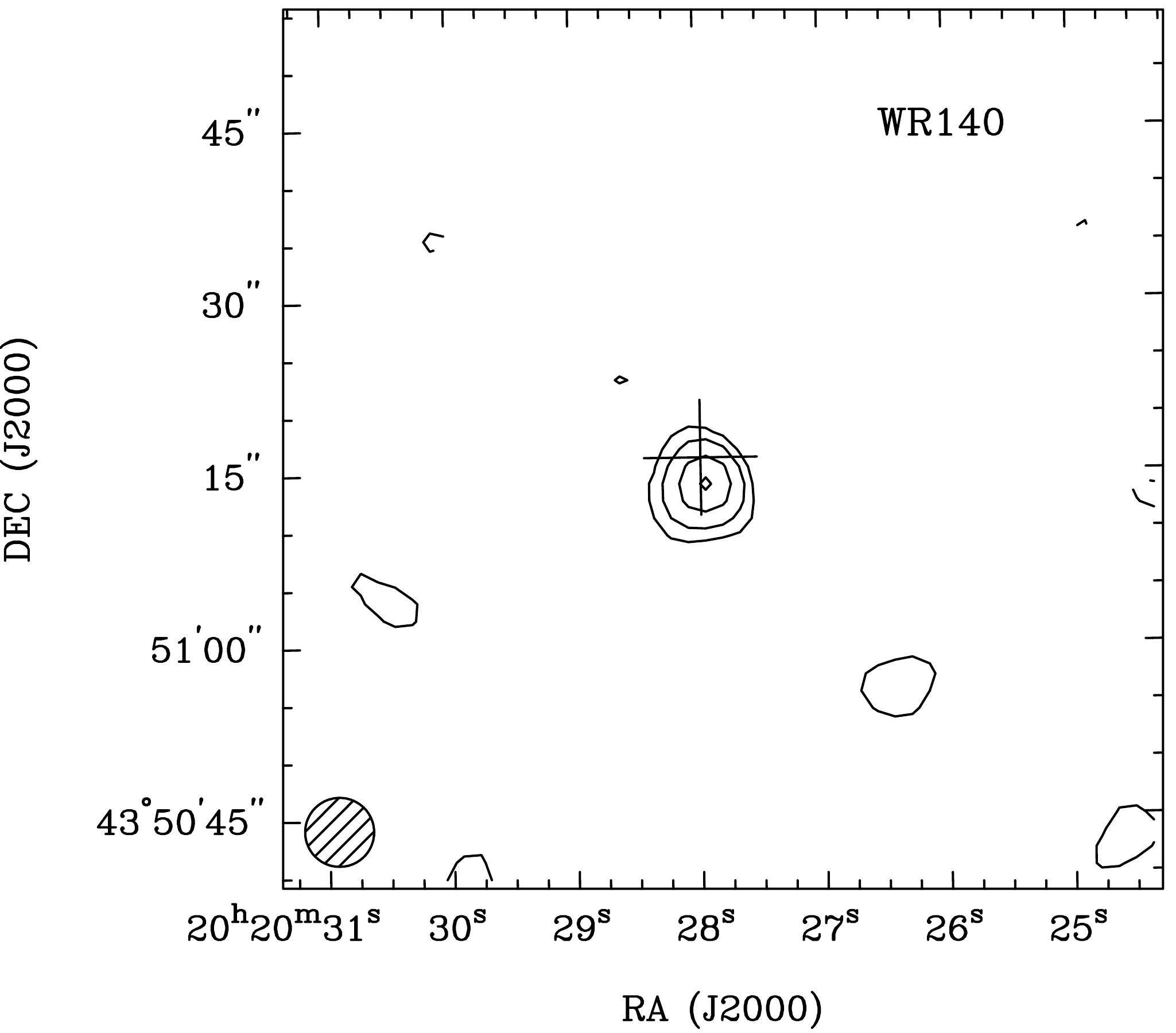}
\caption{GMRT image of the system WR\,140 at 610~MHz. The contour levels are $-$0.16, 0.16  (=2$\sigma$), 0.3, 0.55 and 0.8~mJy~beam$^{-1}$. Hatched, the synthesized beam. {Cross hair: optical position of the system (see Table\,\ref{fluxWR})}.}
\label{Fig2}
\end{center}
\end{figure}

\begin{figure}[!h]
\begin{center}
\includegraphics[width=7cm]{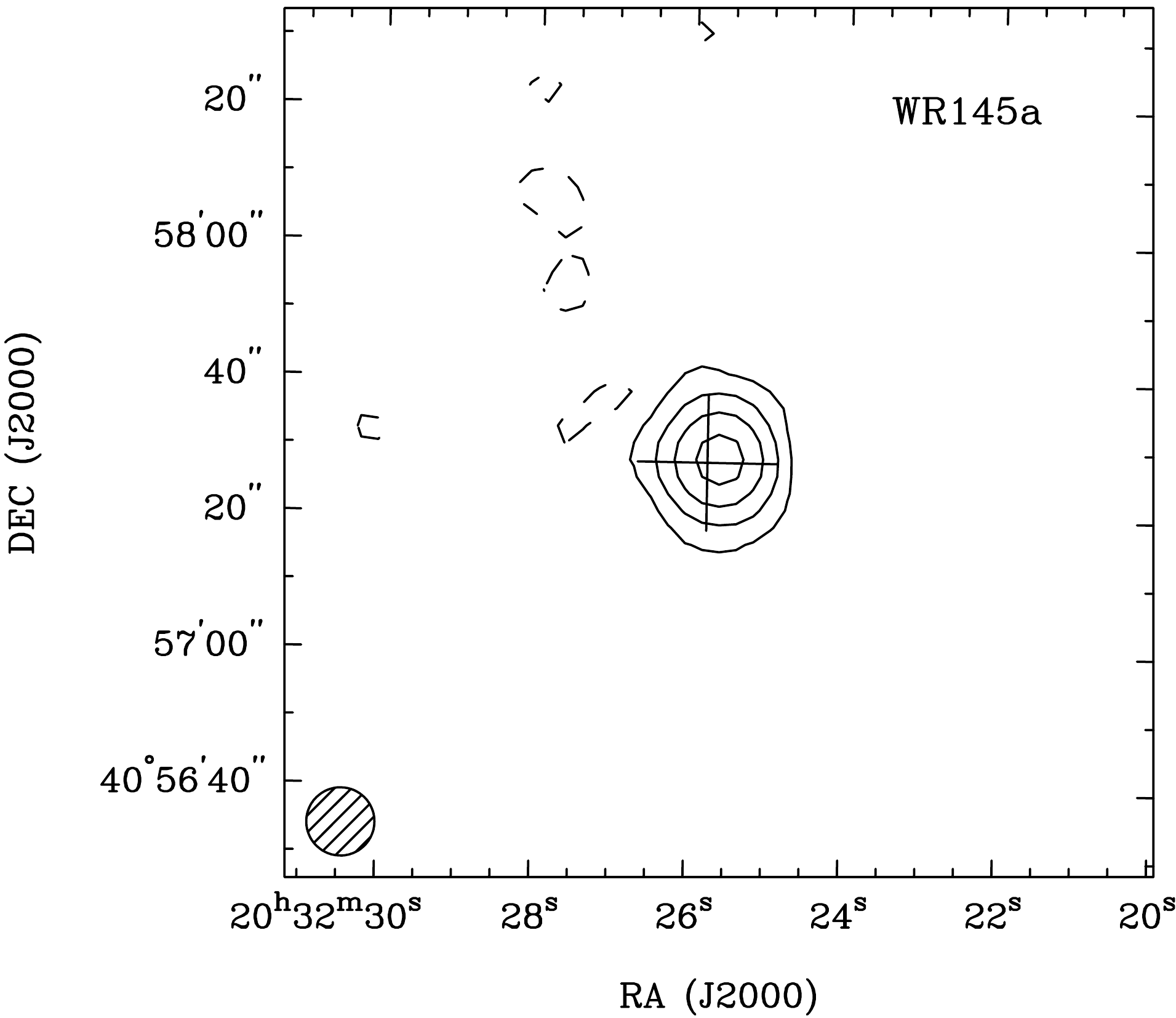}
\includegraphics[width=7cm]{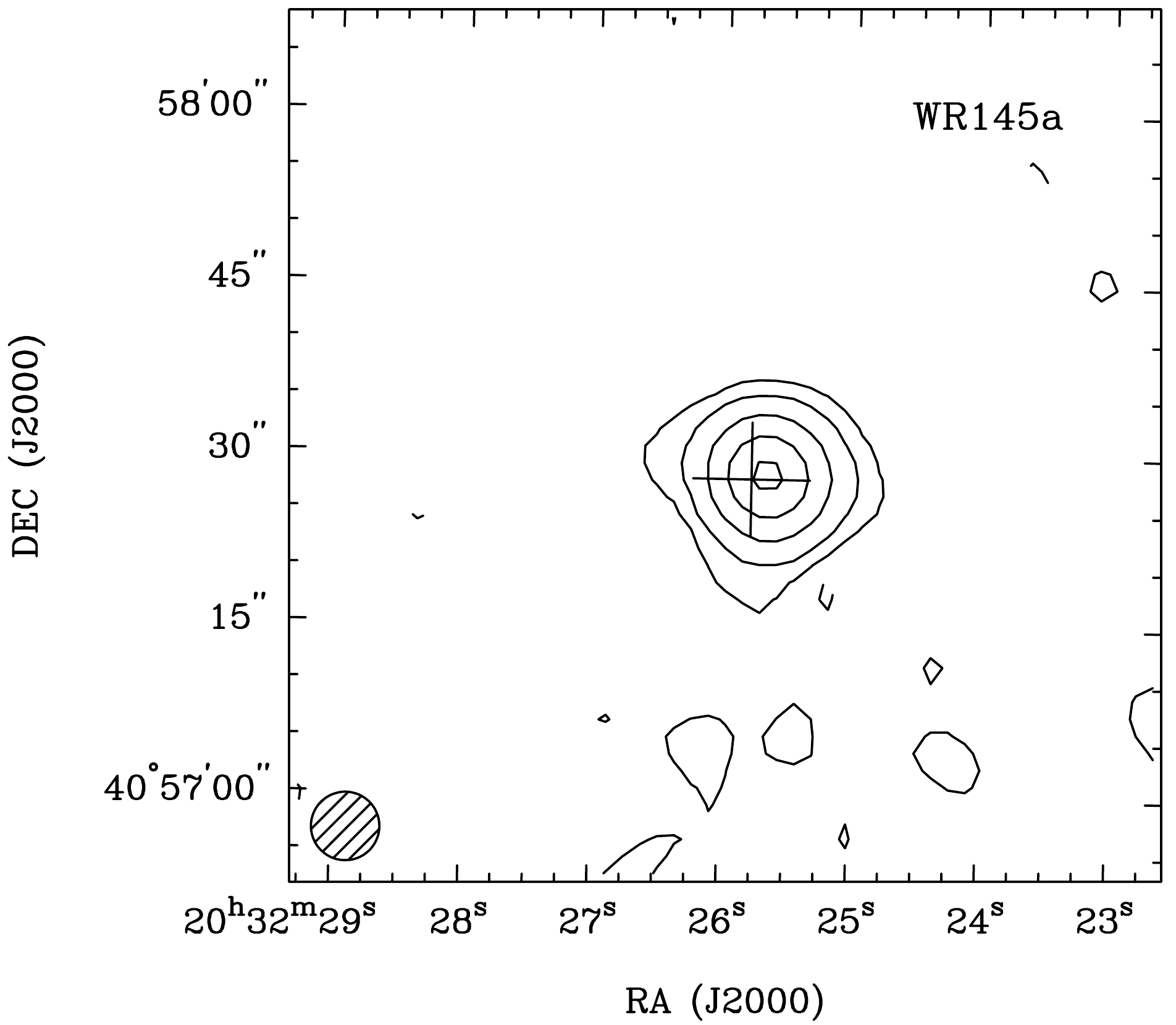}
\caption{GMRT images of the system WR\,145a. Top panel: at 325~MHz; contour levels of  $-$0.8, 0.8 (=3$\sigma$), 4, 10 and 20~mJy~beam$^{-1}$. Bottom panel: at 610~MHz; contour levels of $-$0.3, 0.3 (=2$\sigma$), 2, 10, 30 and 60~mJy~beam$^{-1}$. Hatched, the synthesized beam. { Cross hair: optical position of the system (see Table\,\ref{fluxWR})}.}
\label{Fig3}
\end{center}
\end{figure}

\begin{figure}[!h]
\begin{center}
\includegraphics[width=7cm,angle=-0]{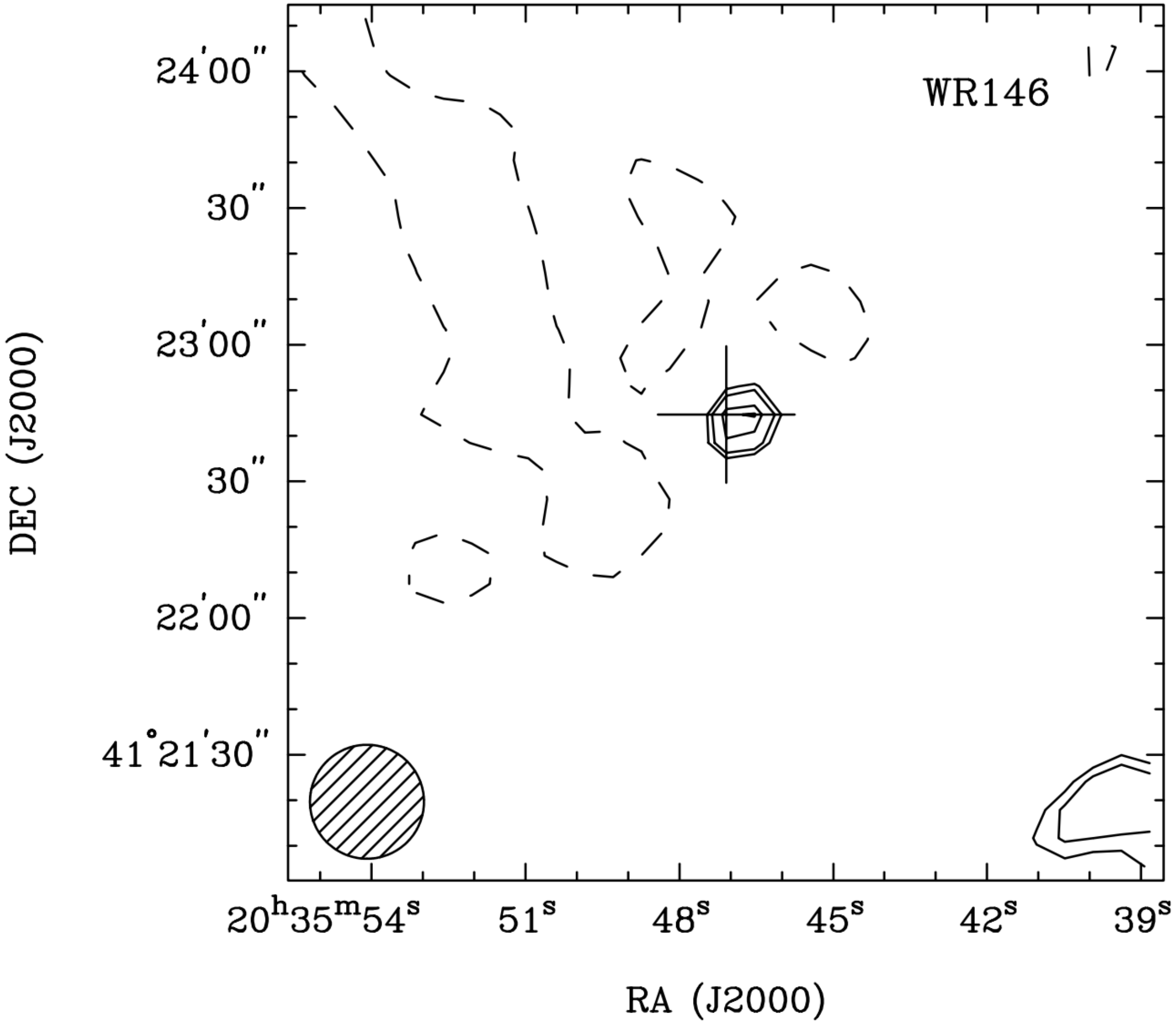}
\includegraphics[width=7cm]{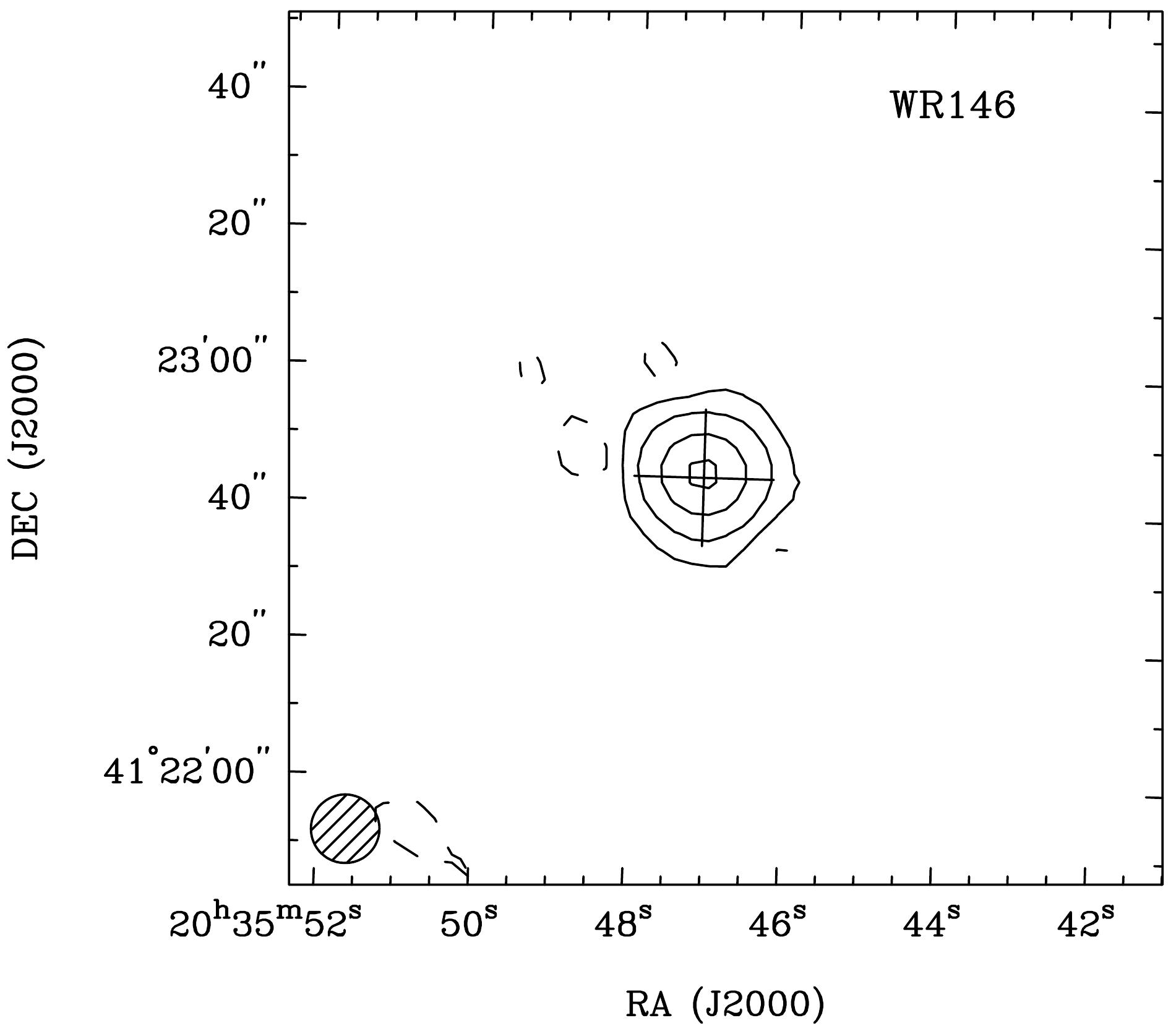}
\includegraphics[width=7cm]{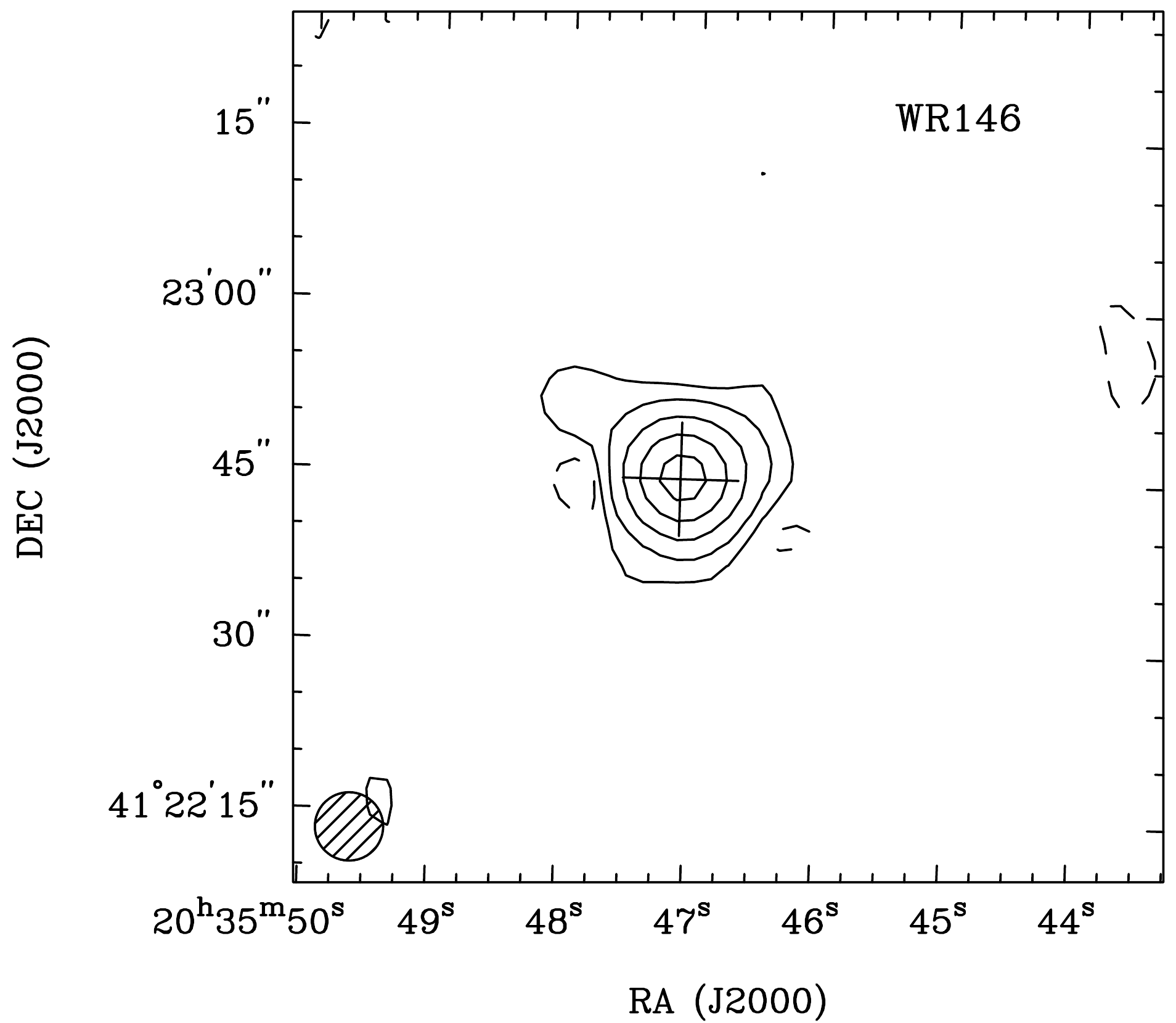}
\caption{GMRT images of the system WR\,146. Top panel: at 150~MHz; contour levels of  $-$18, 18 (=2$\sigma$), 21, 27 and 30~mJy~beam$^{-1}$. Central panel: at 325~MHz; contour levels of  $-$1.5, 1.5 (=3$\sigma$), 10, 40 and 90~mJy~beam$^{-1}$. Bottom panel: contour levels of  $-$0.75, 0.75 (=3$\sigma$), 4, 15 and 40~mJy~beam$^{-1}$. Hatched, the synthesized beam. { Cross hair: optical position of the system (see Table\,\ref{fluxWR})}.}
\label{Fig4}
\end{center}
\end{figure}

\begin{figure}[!h]
\begin{center}
\includegraphics[width=7cm]{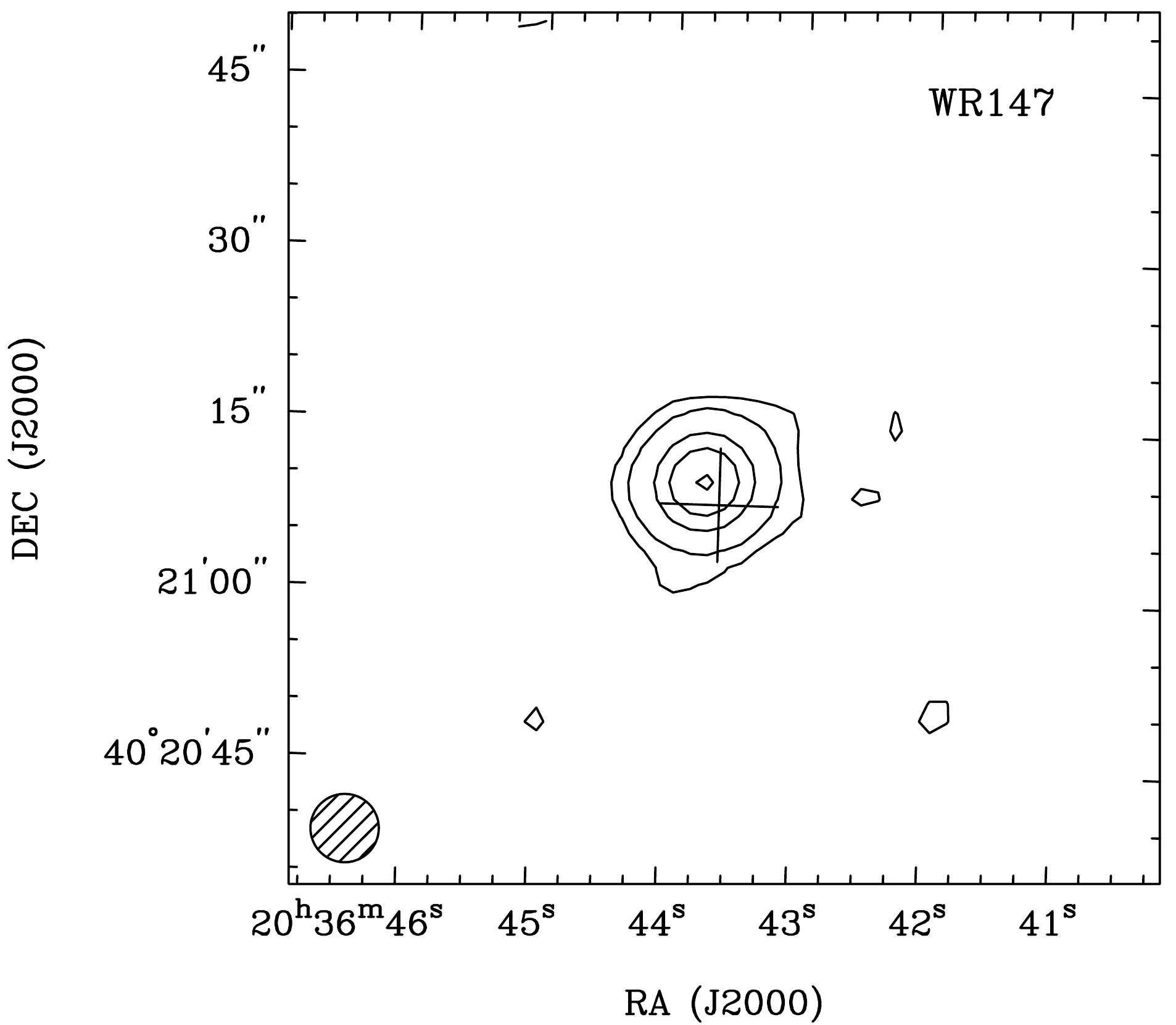}
\caption{GMRT image of the system WR\,147 at 610~MHz. The contour levels are $-$0.24, 0.24 (=3$\sigma$), 2, 5, 10 and 18~mJy~beam$^{-1}$. Hatched, the synthesized beam. { Cross hair: optical position of the system (see Table\,\ref{fluxWR})}.}
\label{Fig5}
\end{center}
\end{figure}

\begin{figure}[!h]
\begin{center}
\includegraphics[width=7cm]{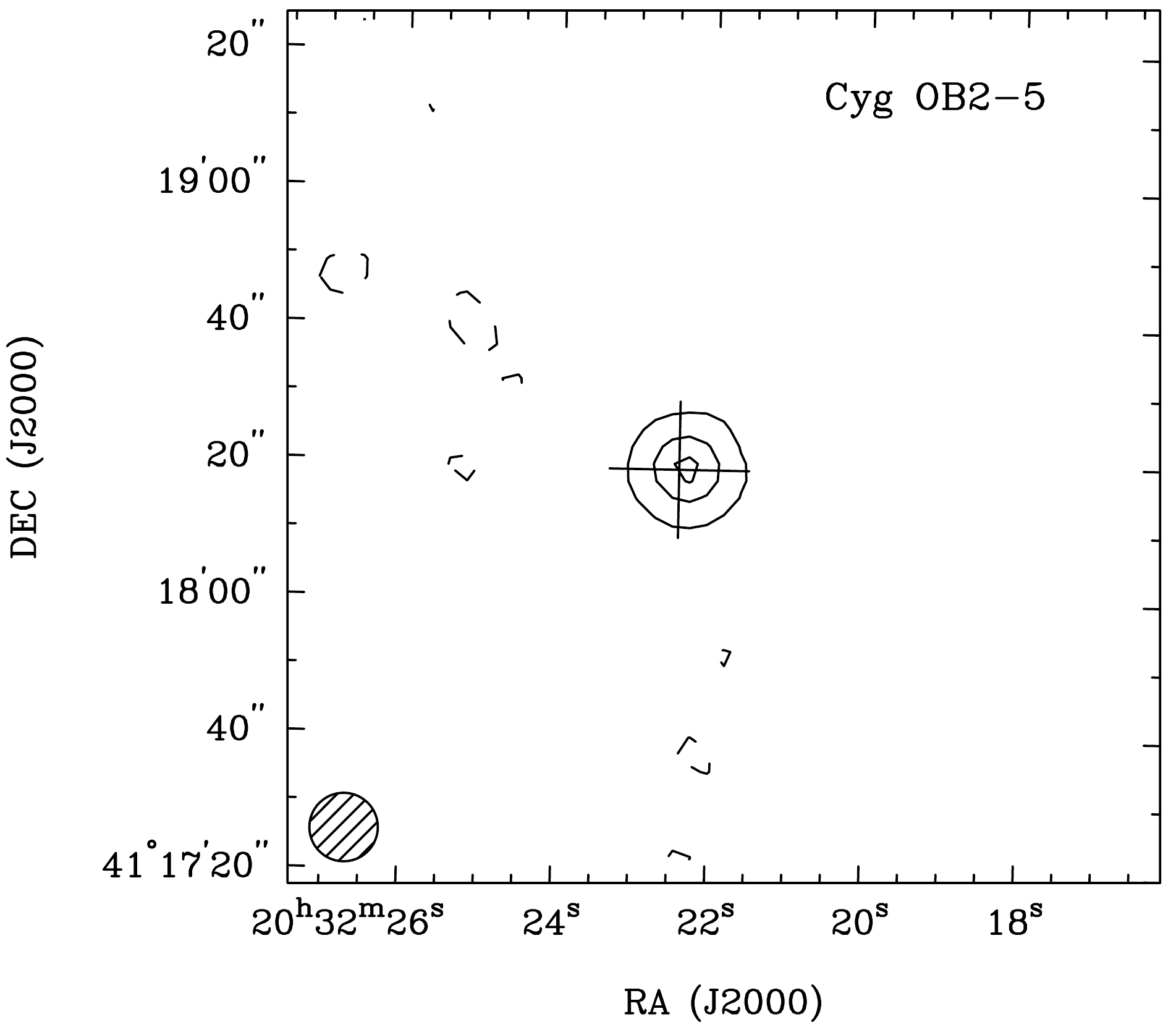}
\includegraphics[width=7cm]{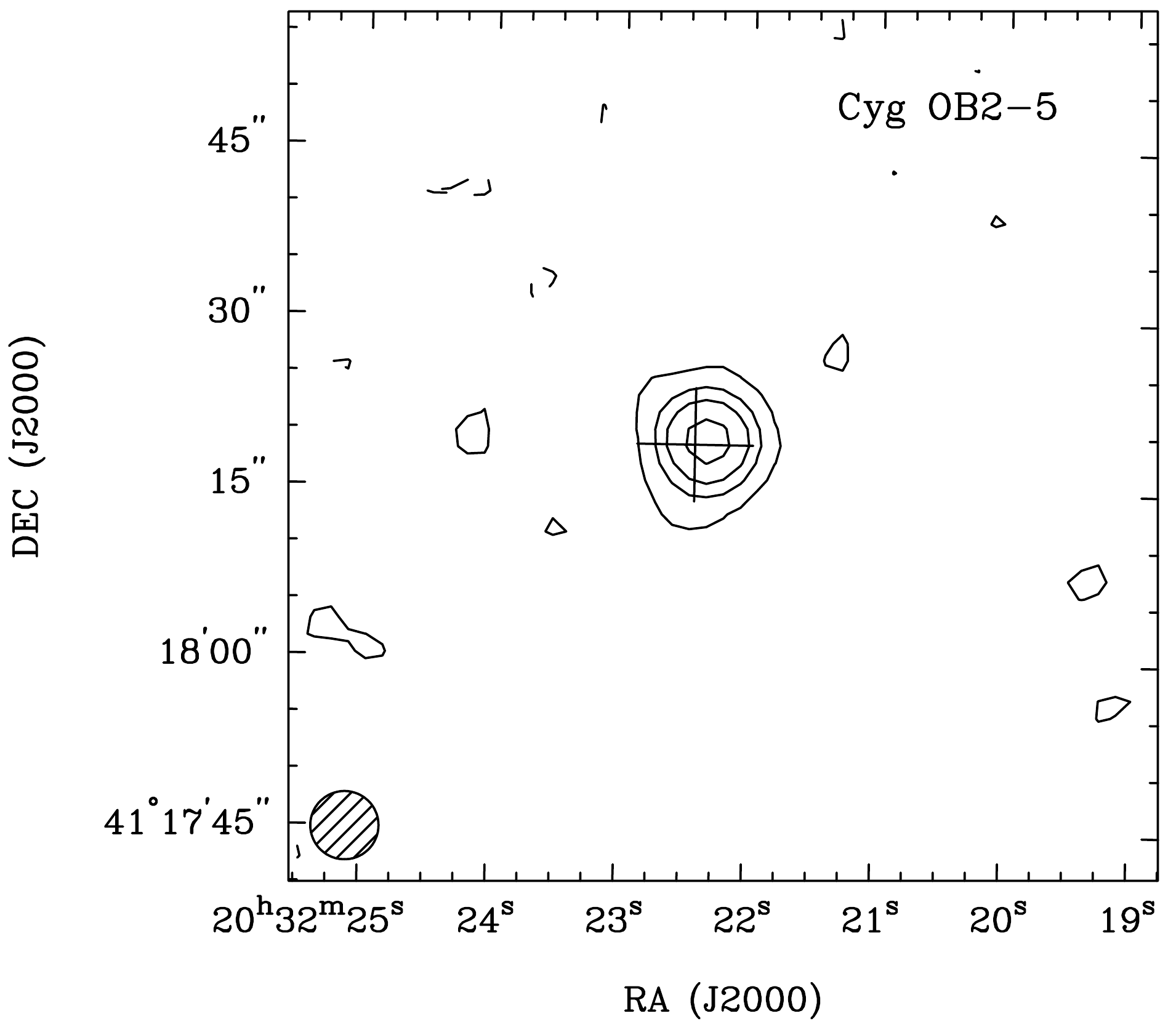}
\caption{GMRT images of the system Cyg~OB2~5. Top panel: at 325~MHz; contour levels of $-$0.7, 0.7 (=3$\sigma$), 3 and 5~mJy~beam$^{-1}$. Bottom panel: at 610~MHz; contour levels of $-$0.15, 0.17 (=2$\sigma$), 0.75, 1.5 and 3~mJy~beam$^{-1}$. Hatched, the synthesized beam. {Cross hair: optical position of the system (see Table\,\ref{fluxO})}.}
\label{Fig6}
\end{center}
\end{figure}

\begin{figure}[!h]
\begin{center}
\includegraphics[width=7cm]{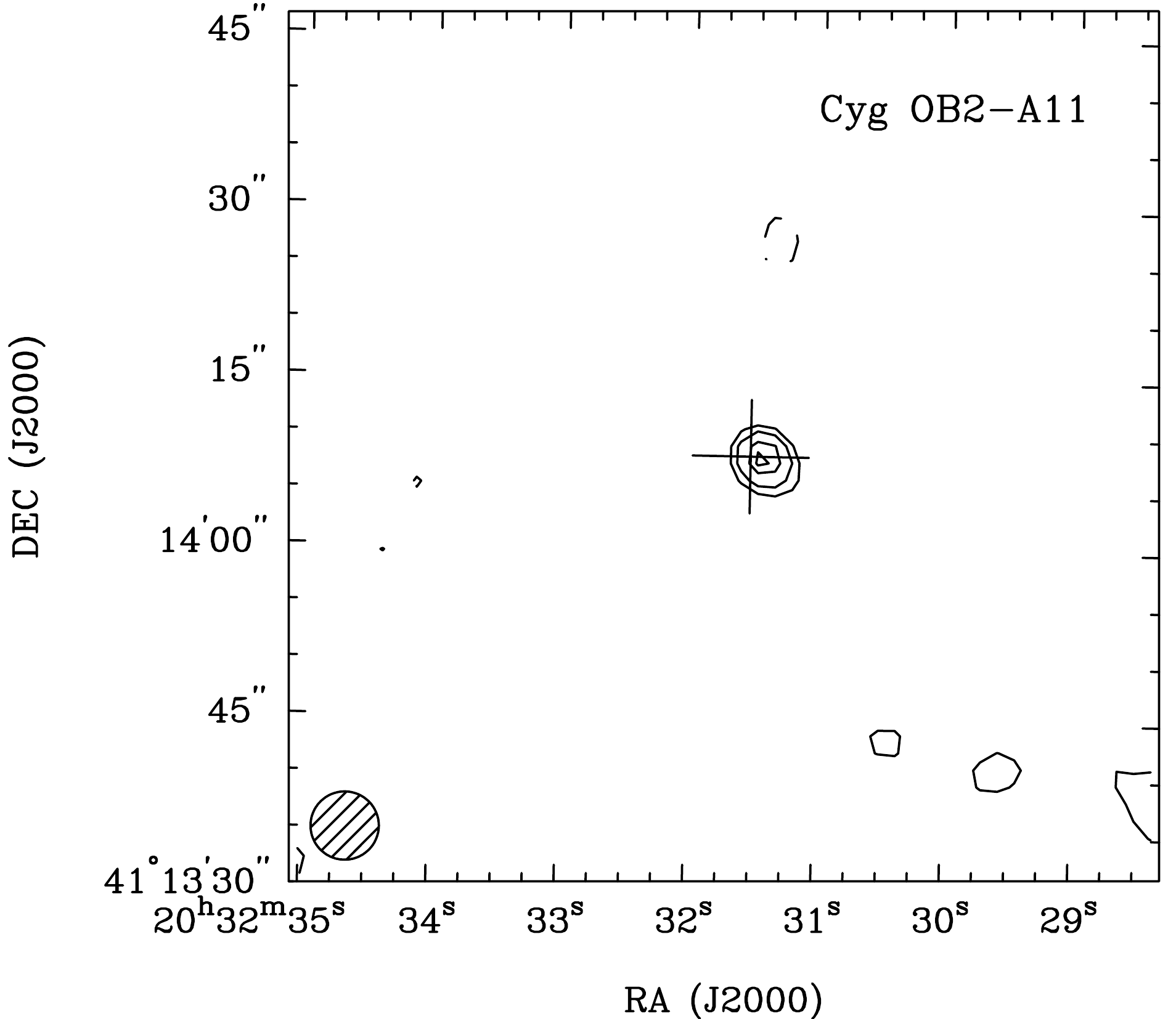}
\caption{GMRT image of the star Cyg\,OB2-A11 
at 610~MHz. The contour levels are $-$0.18, 0.23  (=3$\sigma$), 0.3, 0.4 and 0.45~mJy~beam$^{-1}$. Hatched, the synthesized beam. { Cross hair: optical position of the system (see Table\,\ref{fluxO})}.}
\label{Fig7}
\end{center}
\end{figure}

\begin{figure}[!h]
\begin{center}
\includegraphics[width=7cm]{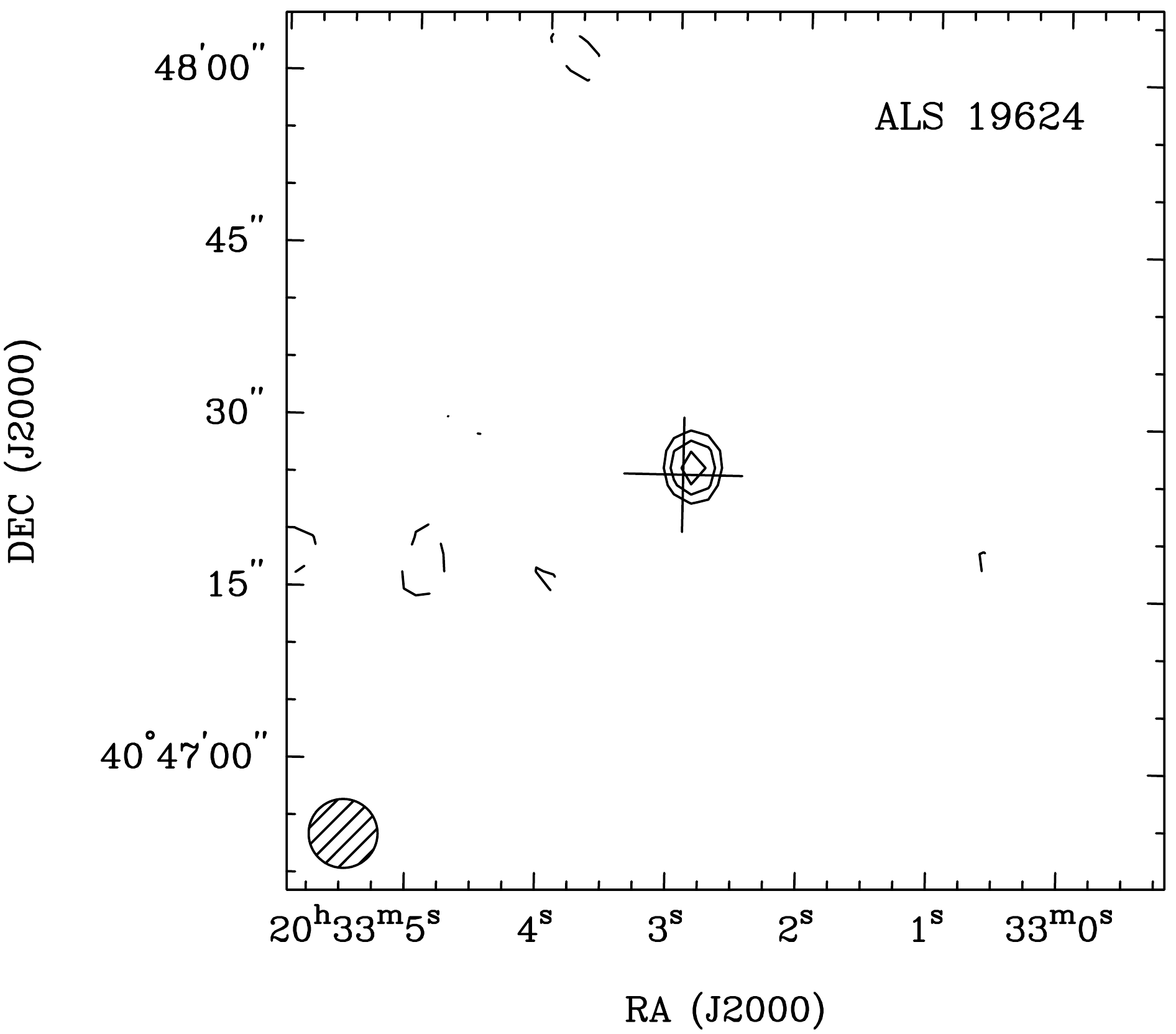}
\caption{GMRT image of the star ALS\,19624 at 610~MHz. The contour levels are $-$0.25, 0.25 (=3$\sigma$), 0.35 and 0.46~mJy~beam$^{-1}$. Hatched, the synthesized beam. { Cross hair: optical position of the system (see Table\,\ref{fluxO})}.}
\label{Fig8}
\end{center}
\end{figure}

\begin{figure}[!h]
\begin{center}
\includegraphics[width=7cm]{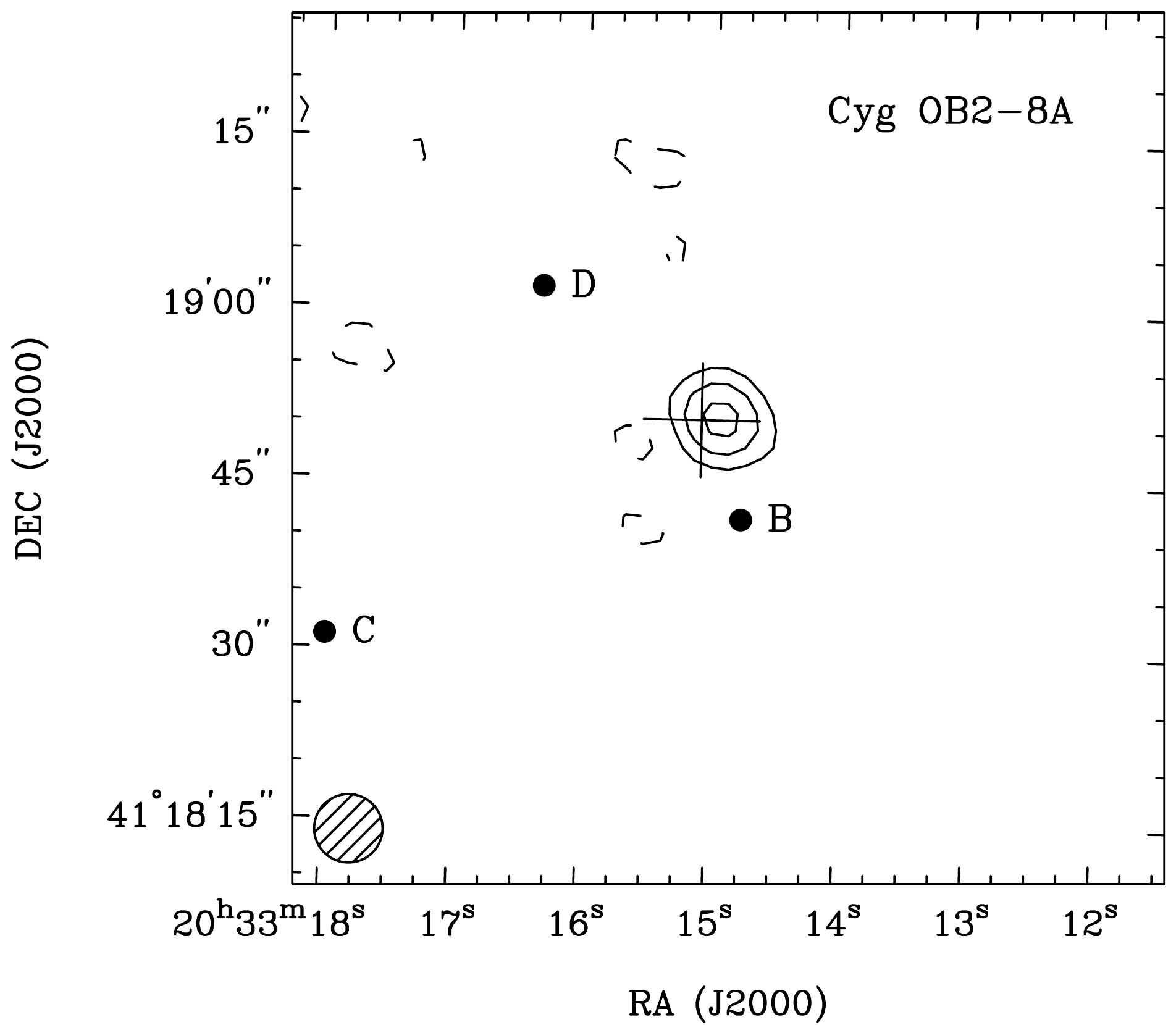}
\caption{GMRT image of the system Cyg\,OB2-8A at 610~MHz. The contour levels are $-$0.35, 0.30 (=3$\sigma$), 1, 2 and 3~mJy~beam$^{-1}$. Hatched, the synthesized beam. The cross hair represents the optical position of the 8A system { (see Table\,\ref{fluxO})}, and the filled circles, those of Cyg\,OB2-8B, -8C, and -8D stars {(see Table\,\ref{nondet})}.}
\label{Fig9}
\end{center}
\end{figure}

\begin{figure}[!h]
\begin{center}
\includegraphics[width=7cm]{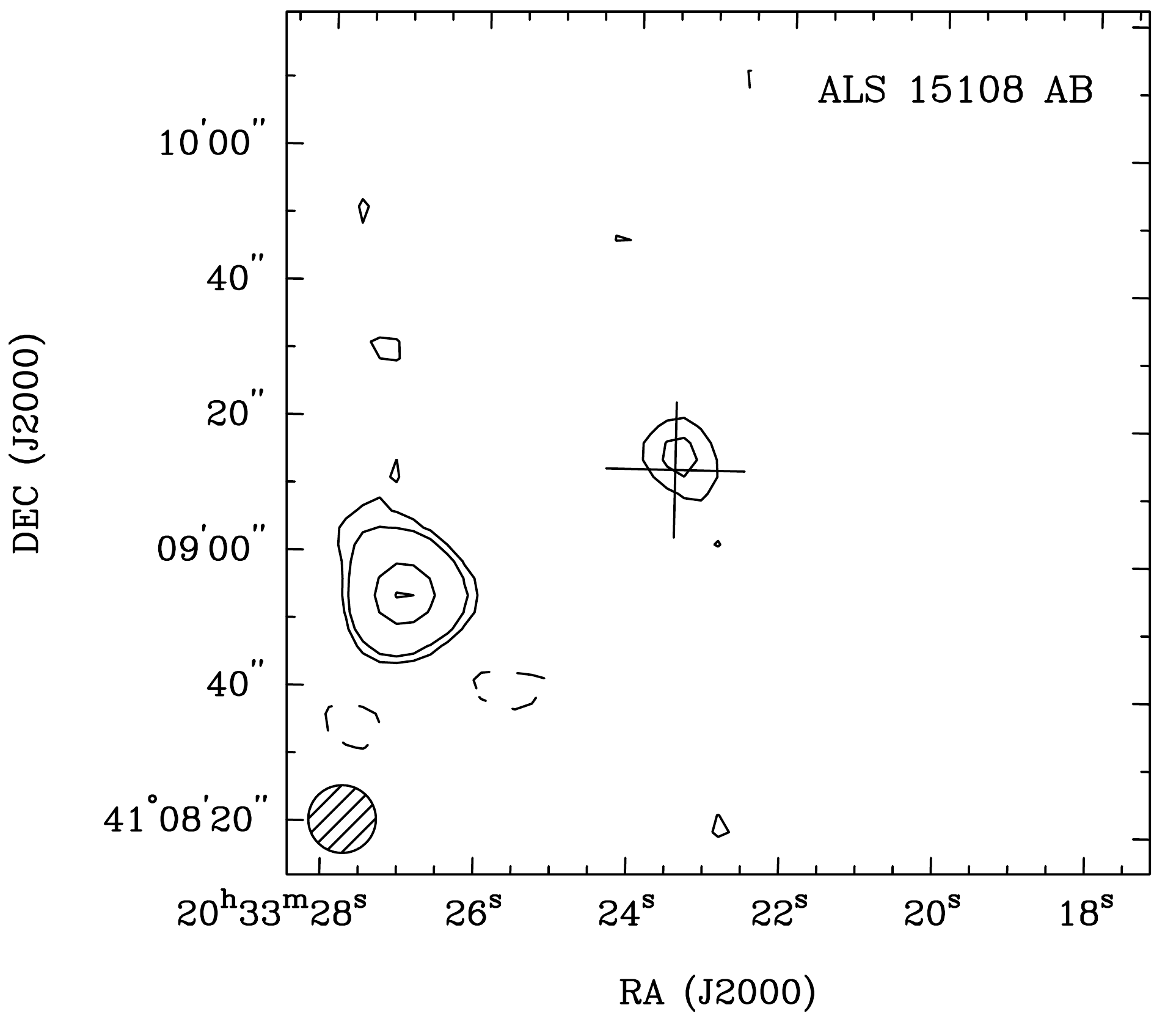}
\includegraphics[width=7cm]{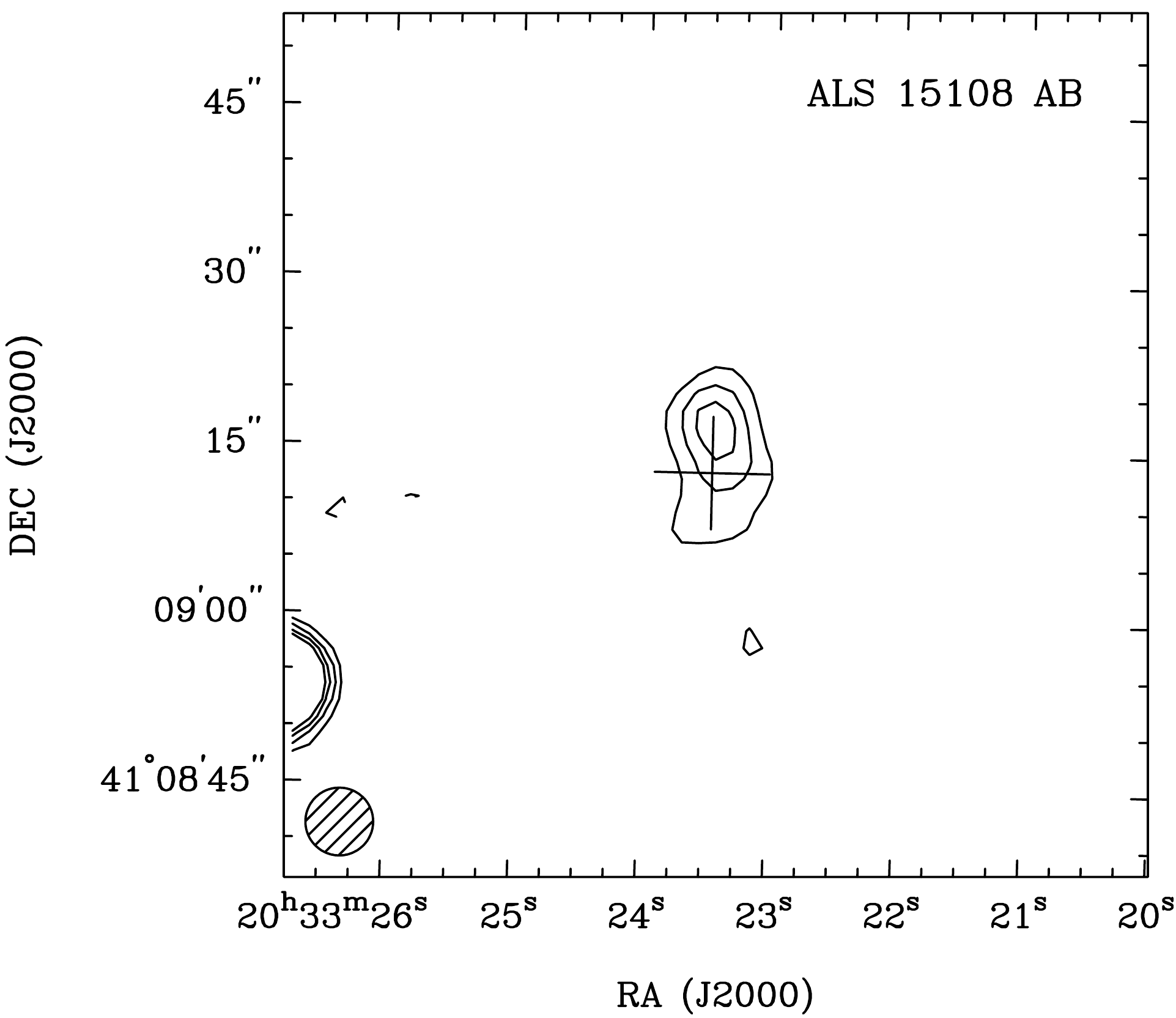}
\caption{GMRT images of the star ALS\,15108~AB. Top panel: at 325~MHz; contour levels of $-$0.5, 0.5 (=3$\sigma$), 1, 5 and 8~mJy~beam$^{-1}$. Bottom panel: at 610~MHz; contour levels of $-$0.25, 0.25 (=3$\sigma$), 0.5, 0.75 and 1~mJy~beam$^{-1}$. Hatched, the synthesized beam. {Cross hair: optical position of the system (see Table\,\ref{fluxO})}.}
\label{Fig10}
\end{center}
\end{figure}

\begin{figure}[!h]
\begin{center}
\includegraphics[width=7cm]{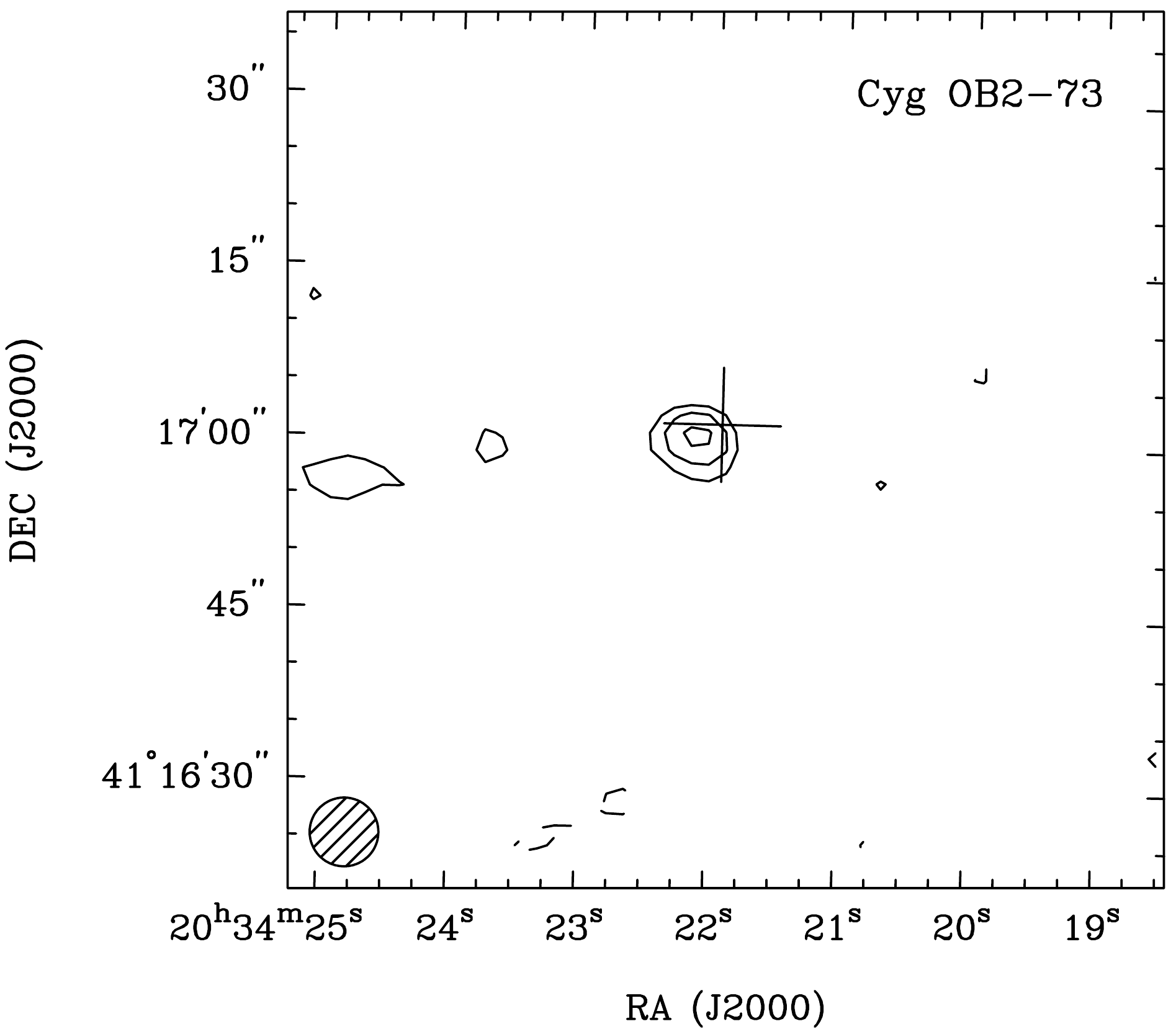}
\caption{GMRT image of the star Cyg\,OB2-73 at 610~MHz. The contour levels are $-$0.22, 0.22 (=3$\sigma$), 3 and 3.9~mJy~beam$^{-1}$. Hatched, the synthesized beam. { Cross hair: optical position of the system (see Table\,\ref{fluxO})}.}
\label{Fig11}
\end{center}
\end{figure}

\begin{figure}[!h]
\begin{center}
\includegraphics[width=7cm]{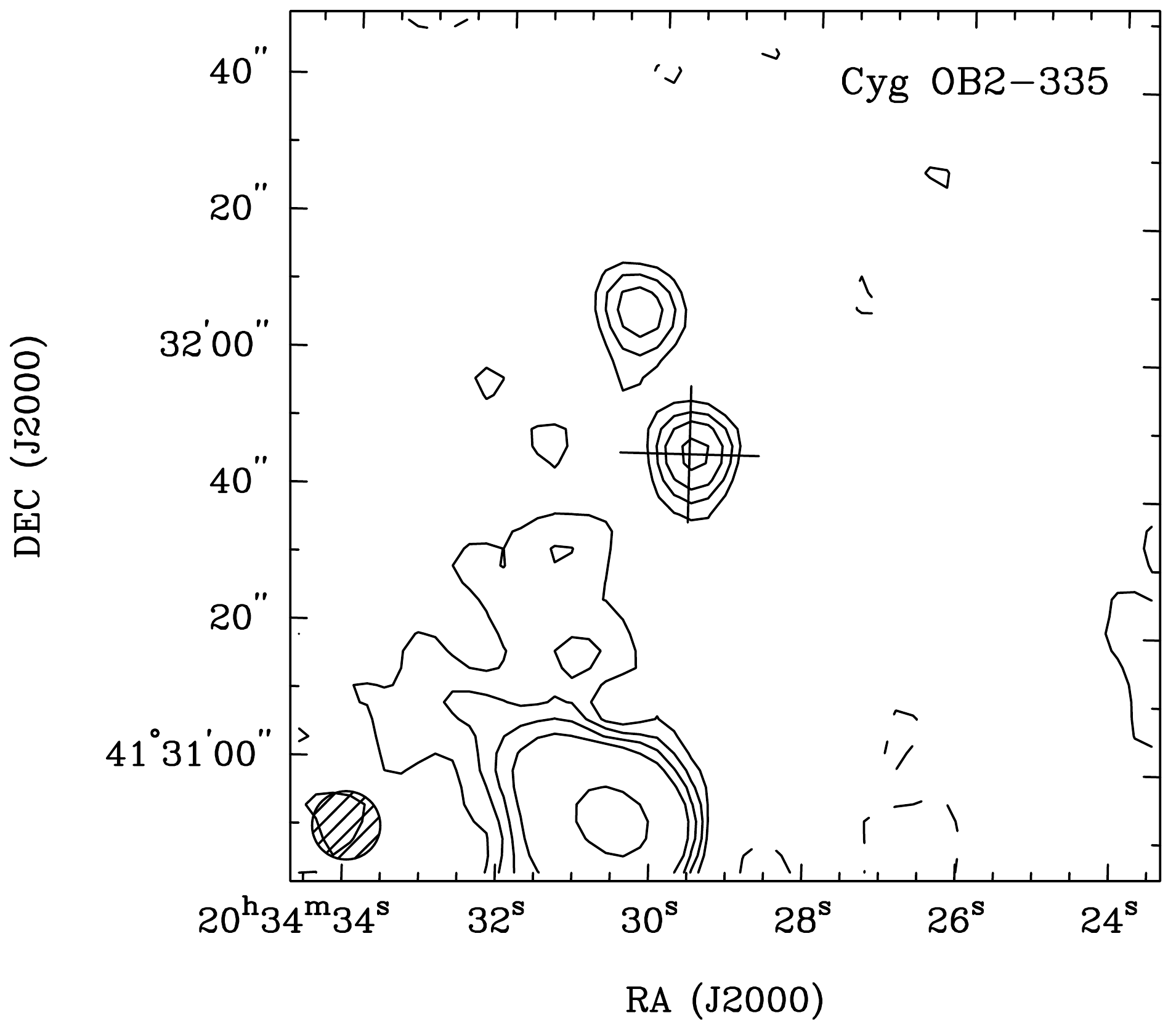}
\includegraphics[width=7cm]{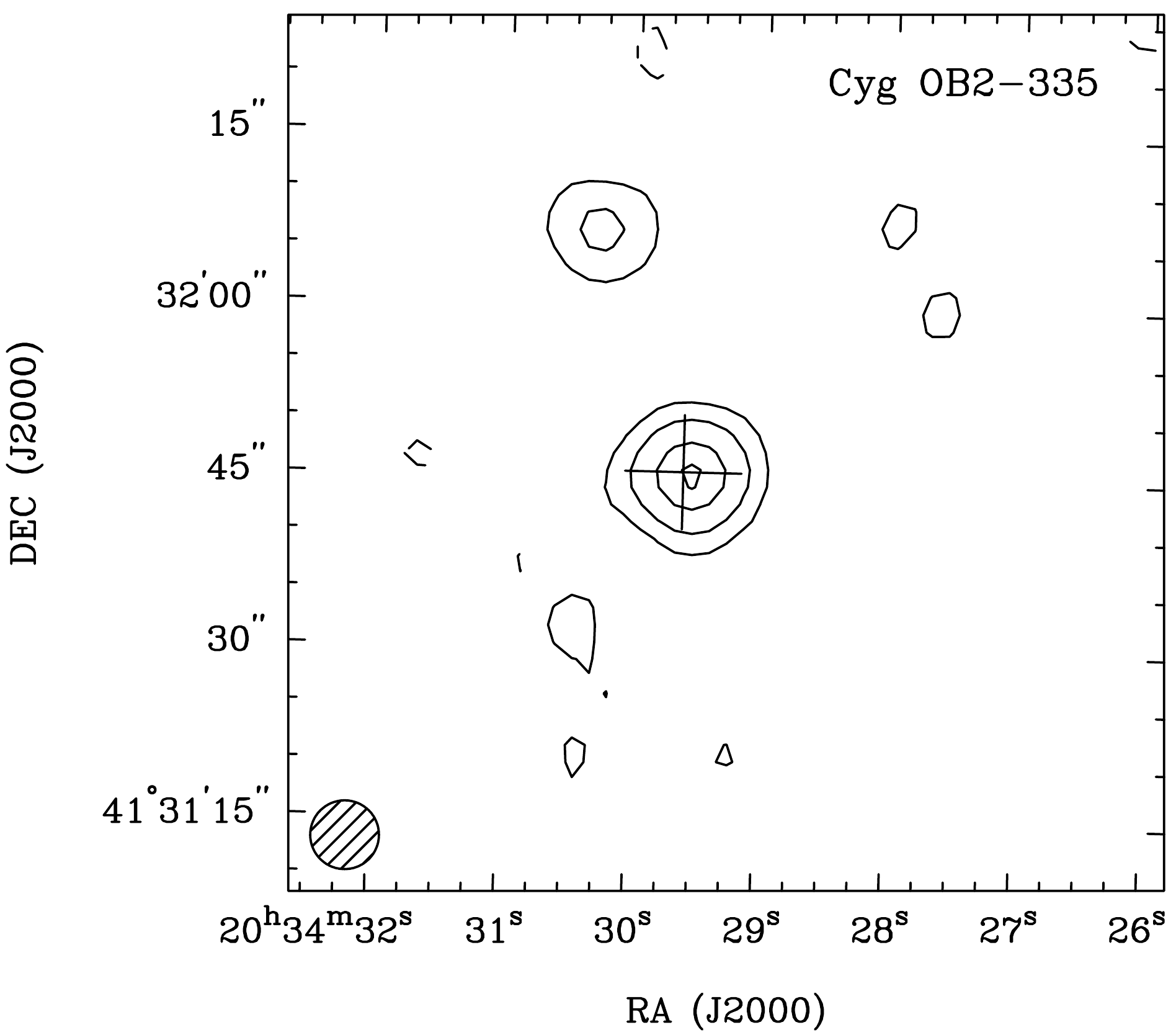}
\caption{GMRT images of the star Cyg\,OB2-335. Top panel: at 325~MHz; contour levels of  $-0.6$, 0.6 (=3$\sigma$), 1, 1.5, 2.3 and 10~mJy~beam$^{-1}$. Bottom panel: at 610~MHz; contour levels of $-0.2$, 0.25 (=3$\sigma$), 1, 3 and 5~mJy~beam$^{-1}$. Hatched, the synthesized beam. { Cross hair: optical position of the system (see Table\,\ref{fluxO})}.}
\label{fig:MWC}
\end{center}
\end{figure}

\begin{figure}[!h]
\begin{center}
\includegraphics[width=7cm]{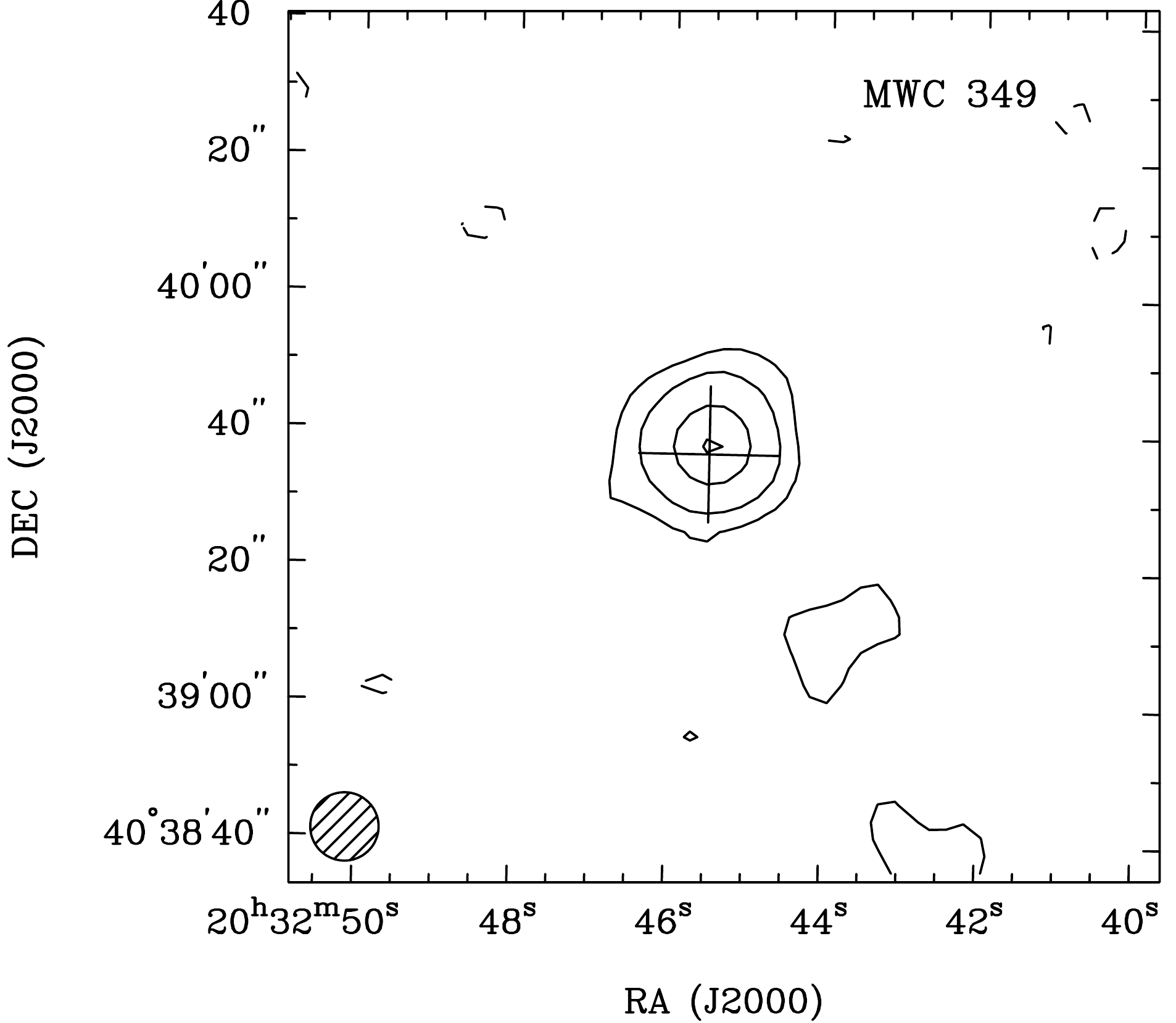}
\includegraphics[width=7cm]{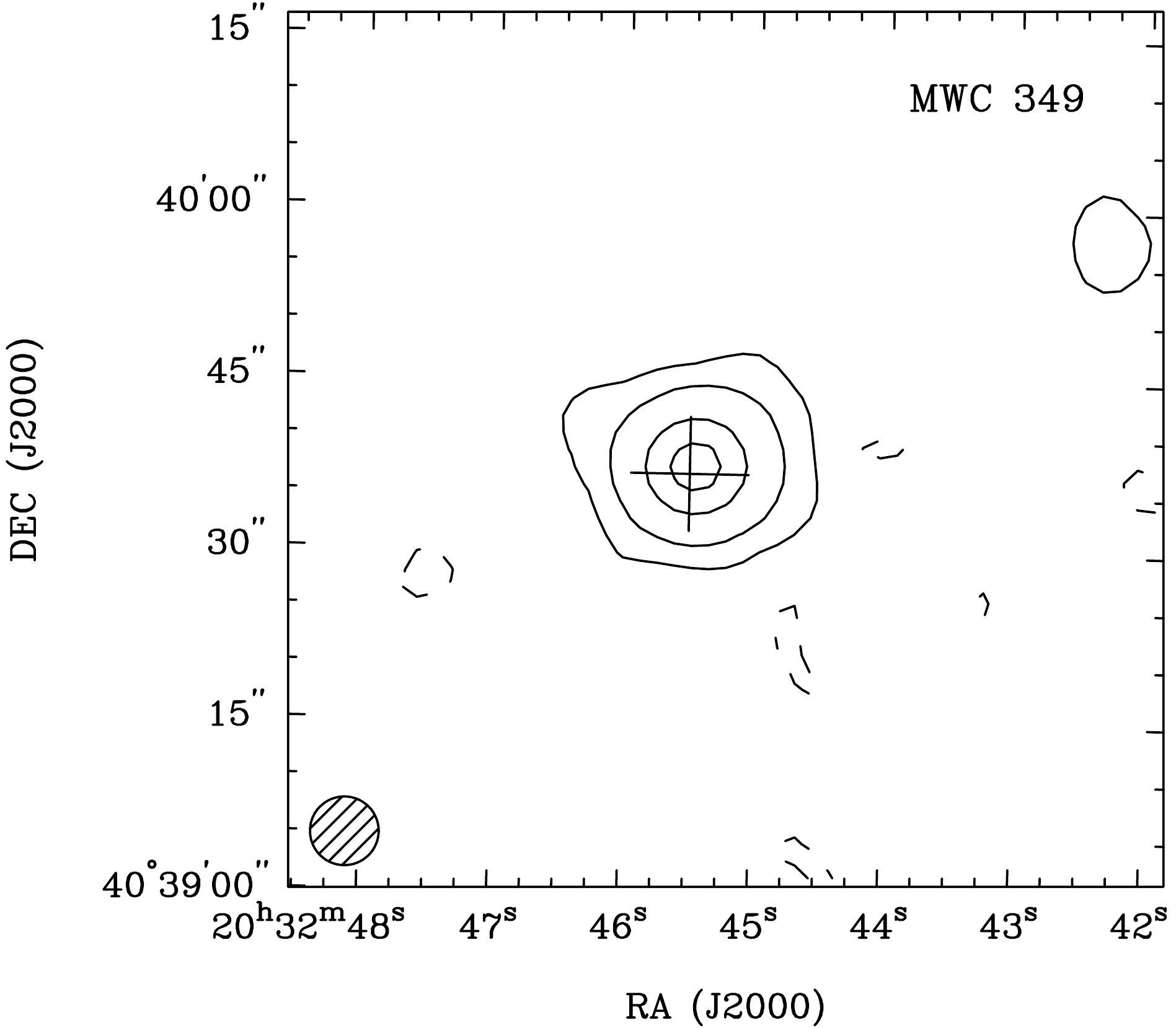}
\caption{GMRT images of the system MWC\,349. Top panel: at 325~MHz; contour levels of  $-$0.7, 0.9 (=3$\sigma$), 3, 10, and 17~mJy~beam$^{-1}$. Bottom panel: at 610~MHz; contour levels of $-$0.3, 0.3 (=3$\sigma$), 2, 10 and 20~mJy~beam$^{-1}$.  Hatched, the synthesized beam. { Cross hair: optical position of the system \citep{UCAC4Cat}}.}
\label{fig:MWC}
\end{center}
\end{figure}

\begin{figure}[!h]
\begin{center}
\includegraphics[width=7cm]{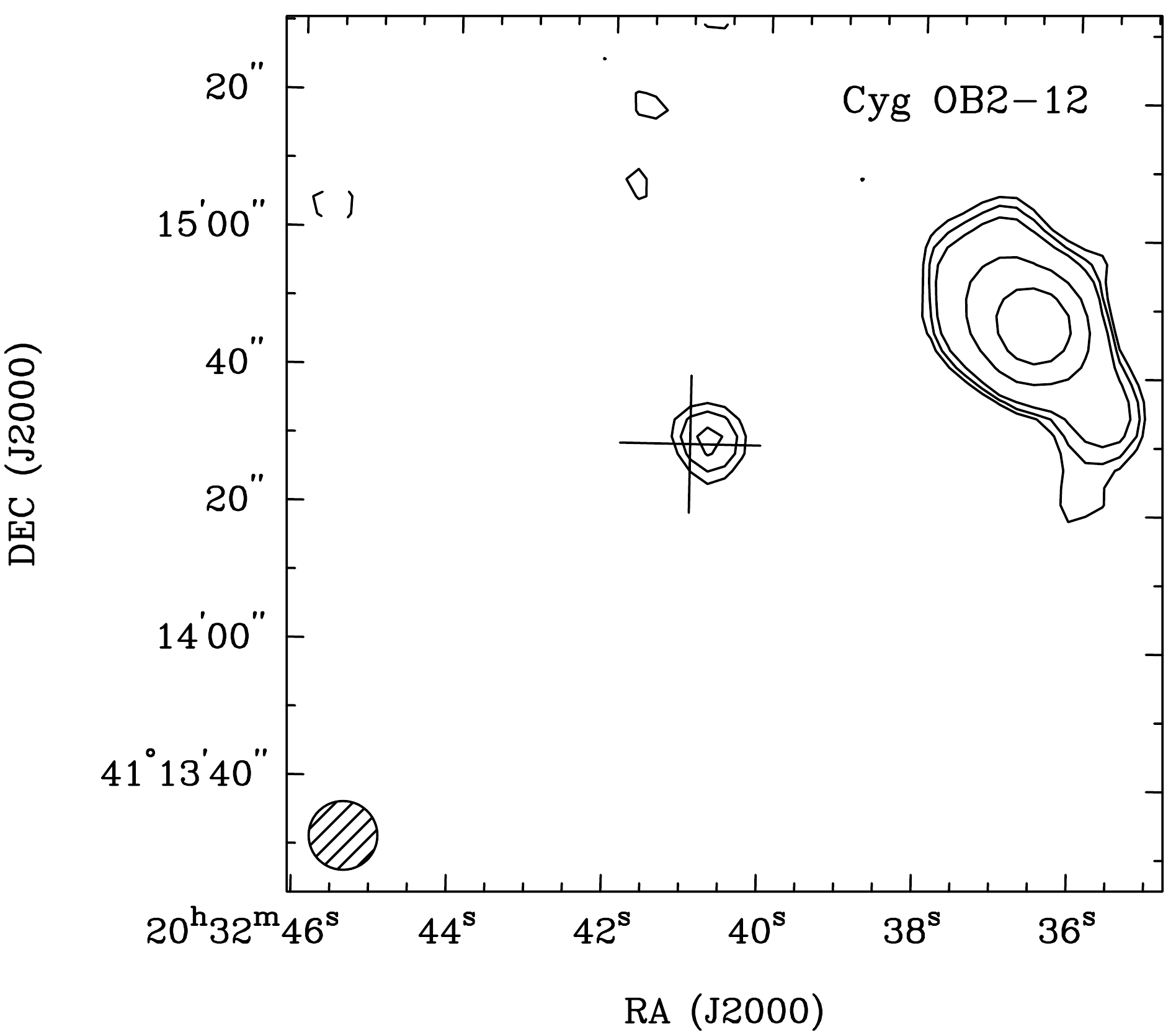}
\includegraphics[width=7cm]{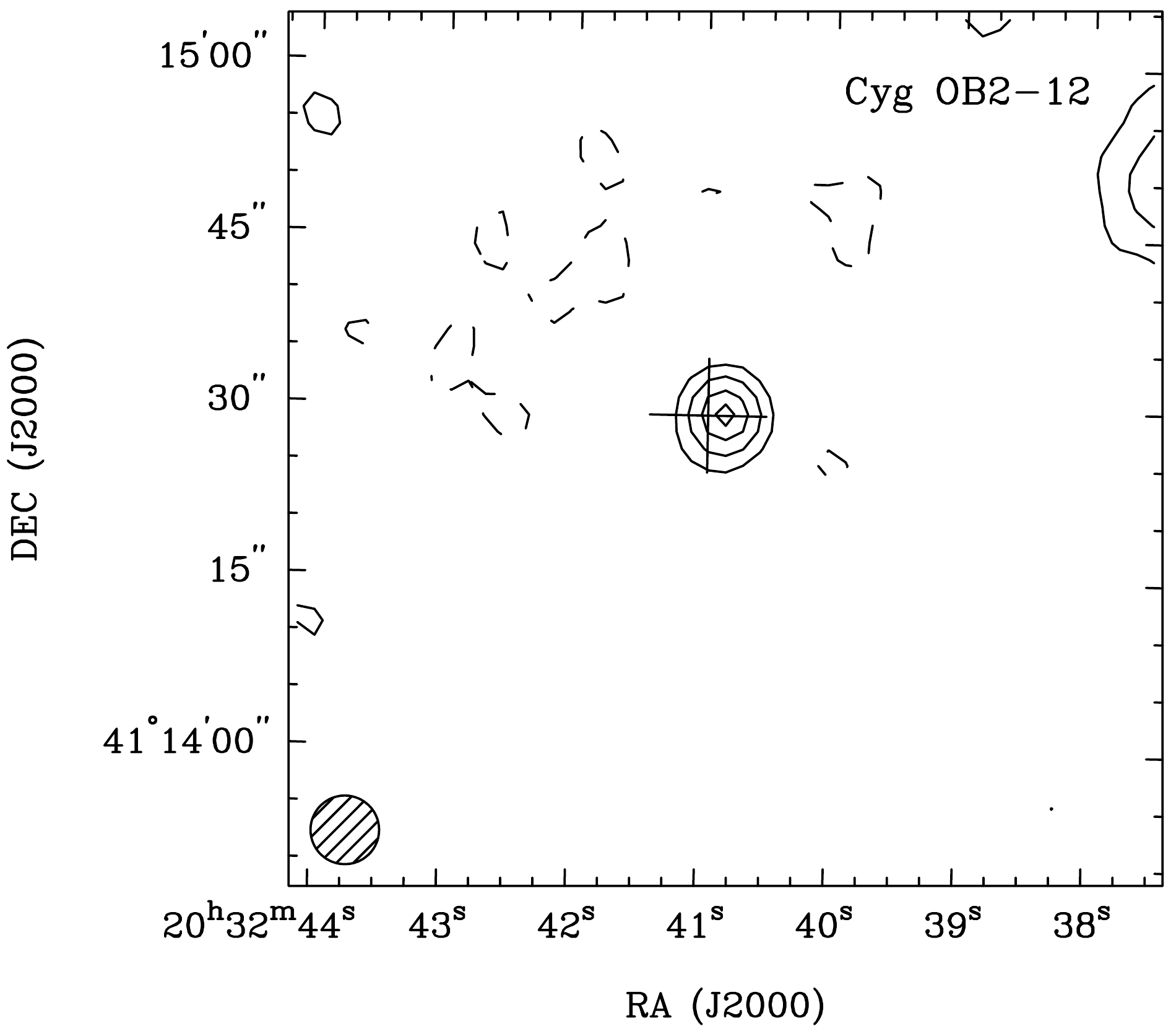}
\caption{GMRT images of the system Cyg\,OB2-12. Top panel: at 325~MHz; contour levels of  $-$0.5, 0.7 (=3$\sigma$), 0.9, 1.2, 3 and 6~mJy~beam$^{-1}$. Bottom panel: at 610~MHz; contour levels of $-$0.2, 0.2 (=3$\sigma$), 0.5, 0.9 and 1.2~mJy~beam$^{-1}$. Hatched, the synthesized beam. { Cross hair: optical position of the system \citep{Gaia2018Cat}}.}
\label{fig:cyg-12}
\end{center}
\end{figure}

\section{DISCUSSION}\label{disc}

\subsection{Detections}

Our analysis led to the detection of eleven WR and O-type stars, either at only one or two MHz frequencies. For them, we derived the spectral indices when possible, and lower limits instead. The main results are summarized in  Tables\,\ref{fluxWR} and \ref{fluxO}.

\subsubsection{The WR systems} 

The WR sample studied here encompasses nine objects,
five of which are known binaries: the four detections reported correspond to that group. 

WR\,140, the emblematic system of the class of PACWBs, is clearly detected 610~MHz. The upper limit at 325~MHz is consistent with a thermal emission component and an absorbed non-thermal component. Though undetermined, the spectral index appears indeed to be positive. 

WR\,145a is one of the brightest WR system in our sample, and this brightness is clearly explained by the physical conditions at work in this system. As a binary system with accreting compact companion, this object does not fit at all in the interpretation scheme of PACWBs at the core of our discussion.

The spectrum of WR\,146 at our measurement frequencies appears to be almost flat, suggesting a composite or turned over nature. 

The results on WR\,147 offer a similar situation as for WR\,140, but with a much higher flux density. 

A more detailed discussion of the systems WR\,140, WR\,146 and WR\,147 is developed in Sect.\,\ref{spect}.

\subsubsection{The detected O-type stars}
The O-type stars sample consists of 87 stars or systems, spanning all spectral type classifications. For many of them, fundamental parameters such as distance are poorly known and we discuss case by case detections. From the seven detected O-type stars, four are confirmed binary systems, one of which presents a non-thermal spectral index between 325 and 610~MHz, while another displays a positive index. Of the three detected O-type stars catalogued as single, one shows a negative spectral index. 

\paragraph{Cyg\,OB2-5.} It is interesting to note that the GMRT measurements clearly confirm the non-thermal nature of the radio emission from this multiple system. If the wind-wind interaction in the 6.7~yr period orbit is at the origin of the synchrotron emission, the stellar separation is large enough to prevent a too severe free-free absorption at these frequencies. 

\paragraph{Cyg\,OB2-A11.} Our results do not point to any detection of this binary system at 325~MHz. The loose upper limit at that frequency precludes any reasonable estimate of the spectral index, leading to a large uncertainty on the nature -- thermal or non-thermal -- of the radio emission.

\paragraph{ALS\,19624.} Same as Cyg\,OB2-A11.

\paragraph{Cyg\,OB2-8A.} The upper limit at 325~MHz and the flux density measured at 610~MHz allow to reject a pure optically thin synchrotron emission at these frequencies. The lower emission level at lower frequencies may be explained either by a pure thermal emission or by a turnover, very likely due to free-free absorption (considering the size of the 22-d orbit and the optical thickness of the winds).

\paragraph{ALS\,15108\,AB.} The negative spectral index reported in Table\,\ref{fluxO} clearly reveals the non-thermal nature of the source. One can therefore claim the identification of a new member in the catalogue of PACWBs. ALS\,15108\,AB is likely a binary system with an astrometric companion at an angular distance of 0.7$''$ \citep{Mason2009}. Assuming a membership to Cyg\,OB2 is established, with a distance of about 1.7\,kpc, this translates into a separation of the order of 1100~AU. Such a stellar separation would indicate a very long orbital period, maybe of the order of centuries. It is not clear whether such a long period system would be able to sustain a wind-wind interaction region capable to significantly accelerate particles. This is especially worth asking given the spectral classification of the star, indicating a rather weak wind in terms of kinetic power. As emphasized by \citet{DeBecker2017}, the class of PACWBs identified so far is mainly populated by objects with stronger winds, even though the detailed requirements for efficient particle acceleration still deserve to be established.
{ Given the above inferences from the reported separation of the known components of the system, the radio detection presented here can be indicative of a third member of the system, closer to one of the stars}.

\paragraph{Cyg\,OB2-73.} Same as ALS\,21079.

\paragraph{Cyg\,OB2-335.} 
The spectral index quoted in Table\,\ref{fluxO}, between { flux density} measurements of 325 and 610~MHz, is positive. In the scenario where the synchrotron emission reported on by \citet{Setia2003} is produced in a wide orbit, we can complement their results, plotting the radio emission from 325~MHz to 8.4~GHz, and explain a { flux density} decay at the lower frequencies advocating for the occurrence of an efficient turnover process. A more detailed discussion of this system is developed in Sect.\,\ref{spect}.

\subsubsection{Comparison with previous results and additional detections} 

The investigation by \cite{Setia2003} 
allowed the detection of three hot massive systems at 350~MHz, namely WR\,145a, WR\,146 and WR\,147. They detected them also at 1.4~GHz, together with Cyg\,OB2-5, Cyg\,OB2-335, and MWC\,349 (a B+B system). The authors also call the attention on other well-known hot  massive stars they could not detect (see their Table\,9). All of them were included in our targets list, except Cyg\,OB2-12, another B-type colliding-wind binary system. We proceeded to check for radio emission at the position of those B-type stellar systems in the GMRT images, and detected both of them.


Thus, complementary to the analysis on WR and O-type stars, we report on the detection of two B-type systems of colliding-wind binaries, which were widely studied in the past, though none are part of the PACWB catalog. Firstly MWC\,349, a B[e]+B0\,III 2.4$''$-wide system. \cite{cohen1985} estimated its distance as 1.2~kpc, according to the spectral type of MWC\,349B\footnote{\citet{strelnitski2013} proposed instead MCW\,349 as a member of the Cyg OB2 association, if A and B components are unrelated.}. This system was detected by \citet{tafoya2004} at 330~MHz with a { flux density} of 30$\pm$10 mJy. We measured the { flux densities} of 28.6$\pm$1.21~mJy (at 325~MHz) and 40.5$\pm$0.54~mJy (at 610~MHz) (Fig.\,\ref{fig:MWC}). The value at 325~MHz reported here resulted in eight times lower noise than the VLA measurement. 
We will analyze the spectrum of MCW\,349 in Sect.\,\ref{spectrum349}.

The second one, {\it Cyg\,OB2-12 (Schulte\,12)}, harbors an evolved B-type hypergiant. It has been confirmed as a binary system thanks to Fine Guidance Sensor interferometric measurements by \citet{FGS2014}, with a stellar separation of about 60~mas, pointing to a likely orbital period of several years. \citet{Marti2007} reported on a detection at 610~MHz with a flux density of 0.93\,$\pm$\,0.22\,mJy. We detected this system at both frequencies (see  Fig.\,\ref{fig:cyg-12}). The peak and integrated { flux density} values at 610~MHz were 1.35$\pm$0.10~mJy and 1.1$\pm$0.15~mJy. At 325~MHz, the corresponding values were 1.3$\pm$0.20 and 1.9$\pm$0.5~mJy, over a higher noise. The error bar on the measurement does not allow to claim any significant variation between both epochs, which would have indicated a likely PACWB status. However, even though the error bars on the flux densities at both frequencies are quite large, we can derive a spectral index $\alpha = -0.87 \pm 0.50$. Such a value, even considering the error bar, cannot be explained by a pure thermal wind emission. In that case, Cyg\,OB2-12 fulfills the criteria to be included in the catalogue of PACWBs. 

A comparison of the results at 325--350~MHz obtained in \cite{Setia2003} and the present work, yields the two common detections of WR\,145a and WR\,146 (WR\,147 was at the field border of the GMRT observations). We detected five more stars; for four of them  (Cyg\,OB2-5, -335, -12, and MWC\,349) \cite{Setia2003} published { flux density} upper limits. The two sets of { flux density} measured values/upper limits are in agreement, with the exception of those of WR\,146. For this system, the value reported here doubles that from the former study. The observations were performed many years apart (1994.5 vs 2013.9). If the difference is due to orbital phase reasons, this makes WR\,146 an unavoidable target to monitor from now on.

\subsection{Non detections} 

\begin{table}[!h]
\caption{Expected thermal {flux densities} at 610~MHz of the WR stars with no previous radio detections. $v_\infty$: terminal velocity. $T_{\rm eff}=0.4 \times T_*$: effective temperature{, where $T_*$ is the stellar surface temperature}.}
\label{expected-fluxes}
\centering
\begin{tabular}{l@{~~~}c@{~~~}r@{~~~}r@{~~~}c}\\
\hline \hline
 Name  & $\dot{M}\times 10^{5}$ & $v_\infty$ & $T_{\rm eff}$ & $S_{\rm 610MHz}^{\rm expected}$ \\
       & (M$_\odot$~yr$^{-1}$) & (km s$^{-1}$) & (kK) & (mJy)   \\
\hline
WR\,138a & 2.0$^{(1)}$ &  700$^{(1)}$ & 16$^{(1)}$ & 0.01 \\ 
WR\,142a & 1.9$^{(2)}$ & 1700$^{(3)}$ & 50$^{(2)}$ & 0.08 \\ 
WR\,142b & 1.6$^{(3)}$ & 1800$^{(3)}$ & 28$^{(3)}$ & 0.06 \\ 
WR\,144  & 2.5$^{(4)}$ & 3500$^{(4)}$ & 44.8$^{(4)}$ & 0.05 \\ 
WR\,145  & 1.6$^{(5)}$ & 1390$^{(5)}$ & 20$^{(3)}$ & 0.11 \\ 
\hline \hline
\end{tabular}
\tabnote{(1): \cite{Gvaramadze2009}; (2): \cite{Pasquali2002}; (3): see average parameters per spectral type classification by \cite{crowther2007}; (4): \cite{Sander2012}; (5): \cite{Muntean2009}.}
\end{table}

\paragraph{WR\,138a, WR\,142a, WR\,142b, WR\,144, WR\,145.} 
Table\,\ref{expected-fluxes} shows the expected { flux densities} due to thermal emission from the stellar wind of these WR stars (main component if a system) not detected at centimeter wavelengths, assuming that the stellar mass-loss rate $\dot{M}$ is known. For all cases, the mean molecular weight ($\mu$), the rms ionic charge ($Z$) and the mean number of electrons per ion ($\gamma$) were adopted as 4, 1, and 1 respectively. 

These undetected systems are either not known to be binary systems (the first four ones), or a known binary with rather short period \citep[WR\,145, 22.5 days,][]{Muntean2009}. In the first case, the lack of a colliding-wind region prevents the particle acceleration mechanism to operate as expected in shocks produced in wind-wind interaction. This makes any synchrotron emission process unlikely. So only the thermal emission of the (stronger) WR wind should be produced, but at low frequencies and at such a distance, the flux density falls clearly below the detection threshold of our measurements: see the expected { flux density} values estimated in Table\,\ref{expected-fluxes} for WR\,138a, WR\,142a and WR\,144, compared to the nominal rms value of the 610-MHz mosaic image. In the second case, provided some relativistic electrons are accelerated, it is very likely that free-free absorption would be strong all along the short orbit,
considering the optical thickness of the WR winds. Here again, no significant synchrotron emission could be expected at those frequencies on top of the thermal emission from the stellar wind(s) which lies below the detection limit of our measurements. Finally, in the case of WR\,138a, tagged as a runaway star, no signs of a bow shock above the noise was detected at any of the observed bands.

\paragraph{Not-detected O-type stars.} The O-type stars without radio emission above 3$\sigma$ ($1\sigma$ = rms) are listed in Table\,\ref{nondet} in the Appendix. 
{ One must take into account that the sample, and any trend derived from it, will be biased by the optical and IR searches performed to discover its stars, due to heavy extinction at those spectral ranges towards the observed region. Besides, the group contains luminosity classes from hyper-giants to dwarfs, thus the origin of their radio emission is expected to be very different. 

Having said so, Table\,\ref{nondet} shows that} the sample is more populated in late-type ($>$\,O7) stars than in earlier objects ($<$\,O7). This is 
in line with expectations for massive star populations. Such populations result in part from the initial mass function favoring lower mass objects \citep[see e.g.][and references therein]{ZY2007}. The stellar distributions are also modulated by the evolution time scale, longer for later-type stars. This { behavior should} be explained by the { flux density}-limited nature of the census of stars. Later type objects are overall fainter, and observational biases lead to significant underestimates of their number.

The lack of detection in general can be interpreted in line with a few basic ideas. If they are not synchrotron radio emitters, the thermal emission from their winds, at a distance of a few kpc, should not be bright enough to be detected at the sensitivity of our measurements. In addition, let us keep in mind that thermal emission is characterized by a positive spectral index, thus much fainter at lower frequencies. Second, an additional radio emission component of synchrotron origin is expected only if these objects are at least binaries. A significant fraction of them should be binaries, but the multiplicity of many of them still needs to be established. Finally, if synchrotron emission is active in some of these objects, free-free absorption is likely to be at work, and will be especially efficient in systems with periods not longer than a few weeks. Consequently, the expected rise of the radio emission at lower frequencies may be compensated by a significant attenuation.

In line with discussions in \citet{DeBecker2017}, late-type objects are underrepresented in the catalogue of PACWBs. This comes from the lower kinetic power available in weaker wind systems, resulting in less available energy to feed non-thermal processes in the colliding-wind region. As a result, their potential weak synchrotron radio emission would lay below the sensitivity limit of our observations. The bulk of the { sampled} population is thus located in the less favorable part of the stellar parameter space to display hints for particle acceleration. For the few earlier systems, as discussed above, the lack of detection may either be attributed to a single star status or a significantly free-free absorbed synchrotron emission. We also clarify that a relevant approach would consist in checking for radio detections among actual binaries, in order to address the issue of the fraction of PACWBs among CWBs, as discussed by \citet{DeBecker2017}. This would require a good knowledge of the multiplicity status of all objects in the field, but this information is poorly determined.

Among the most studied O-type systems in Cyg\,OB2, there is {\it Cyg\,OB2\,9 (Schulte\,9)}. This 2.35-yr period system, made of O5--O5.5I and 3--O4III components, is a PACWB whose radio emission at GHz frequencies was investigated in detail by \cite{Blomme2013}. It was detected at 610\,MHz by \citet{Marti2007} with a flux density of 1.24\,$\pm$\,0.20\,mJy, though not detected in our more recent data. The { flux density} value at the position of the system, in the 610~MHz images presented here, is 0.2~mJy beam$^{-1}$, at the level of the area rms (we note that diffuse emission is present in the surroundings). This may indicate a variation in the radio emission, indicative of non-thermal emission in line with the criteria proposed by \citet{catapacwb} for synchrotron radiation produced in the colliding-wind region.

\subsection{On the radio spectra characteristics}\label{spect}

Many of the systems under study have been detected at other radio bands, and the measured {flux densities} reported here allow to complement their spectra. Some of them clearly show a composite spectrum, with thermal and non-thermal contributions. 

The radio emission from colliding-wind binaries is well described by the calculations of \citet{doughertymodel2003} and \citet{pittardmodel2006}. The authors produced detailed models of the spectral and spatial distribution using hydrodynamical simulations and solving the radiative transfer equation, and applied them to the systems WR\,140 and WR\,147. Although those systems were widely observed at radio waves, some parameters, as diverse as inclination or wind clumping factor for instance, remain unknown.

The mentioned models of a CWB take into account thermal (free-free) emission from the individual winds of the hot massive stars that compose the system, and the non-thermal (synchrotron) radiation from the CWR, prone to be modified by thermal absorption along the spectrum. 
If $\alpha_{\rm ff}$ and $\alpha_{\rm NT}$ are the thermal and synchrotron spectral indices, and $\tau_0$ is the optical depth at 1~MHz \citep[see for instance][]{Setia2000}, the measured { flux density} $S_\nu$ can be expressed as

\begin{equation}
    S_\nu = {\rm A}\,\nu^{\alpha_{\rm ff}}\,\, + {\rm B}\,\nu^{\alpha_{\rm NT}}\,\exp^{-\tau_0\,\nu^{-2.1}},\label{eq:total-flux}
\end{equation}

\noindent where A and B are constants, { and the model assumes that the attenuating ionised medium is external to the population of relativistic electrons that are producing the non-thermal emission.}

{ At low} frequencies, the absorption processes of synchrotron self-absorption (SSA) and the Razin-Tsytovich effect \citep[RTe, see][and references therein]{white-chen1995} are expected. \citet{williams1963} showed that for a source size of $\theta$ arcsecs in a magnetic field of $M$ Gauss, SSA is critical at the frequency

\begin{equation}
    \nu_{\rm SSA} \approx 2.145 \, \left( {\frac{S_{\nu_{\rm SSA}}}{\theta^2} } \right)^{2/5}\,M^{1/5} \,\,\,\, {\rm MHz},
\label{eq:freqSSA}
\end{equation}
 
\noindent if  $S_{\nu_{\rm SSA}}$ is the maximum flux density in mJy. And the cutoff frequency for RTe \citep{pacholczyk197}, if the plasma density is $n_e$, can be approximated by

\begin{equation}
  \nu_{\rm RTe} = 20\, n_e \,/\,M \,\,\,\, {\rm Hz}.
\label{eq:freqRTe}
\end{equation}

In this study we provide { flux density} values of various { colliding-wind} systems, at MHz frequencies. It is not the intention here to dwell into the hydrodynamic models, but performing a phenomenological fit, with illustrative purposes, to exploit the information at very low frequencies and the turnover processes mentioned above: in some way, following the approach of \citet{Setia2001}. The exercise allows to limit parameters like the non-thermal spectral index and the relevancy of FFA. We discuss individually here the binary systems WR\,140, WR\,146, WR\,147, Cyg\,OB2-335 and  MWC\,349. We left aside (i) the multiple system Cyg\,OB2-5, as complex geometries and system structure, more than one colliding-wind region, would increase the degrees of freedom and the putative scenarios to consider, (ii) the 22d-period binary Cyg\,OB2-8A, since the detection at 610~MHz was obtained by averaging observations one month apart, blurring any indication of the phase-locked emission that was discovered by \citet{Blomme2010}, 
and (iii) Cyg OB2-12, because the measured { flux densities} presented considerable uncertainty. 

The fits were done through a python routine including the function optimize.curve\_fit, that uses the Levenberg-Marquardt algorithm\footnote{https://docs.scipy.org/doc/scipy/reference/generated/scipy. optimize.curve\_fit.html}, { \citep[see also][]{scypy}}. The routine finds optimal values for the parameters so that the sum of the squared error of the difference (fitted-function value -- measured { flux density}) is minimized.
The errors in { flux density}  values are used as weights in the least-squares problem.
{ A caveat here: except for WR\,140, we will fit data of different orbital phases, which will most impact on short-period systems.}

\subsubsection{WR\,140} 
\begin{figure}[!t]
\begin{center}
\includegraphics[width=8cm]{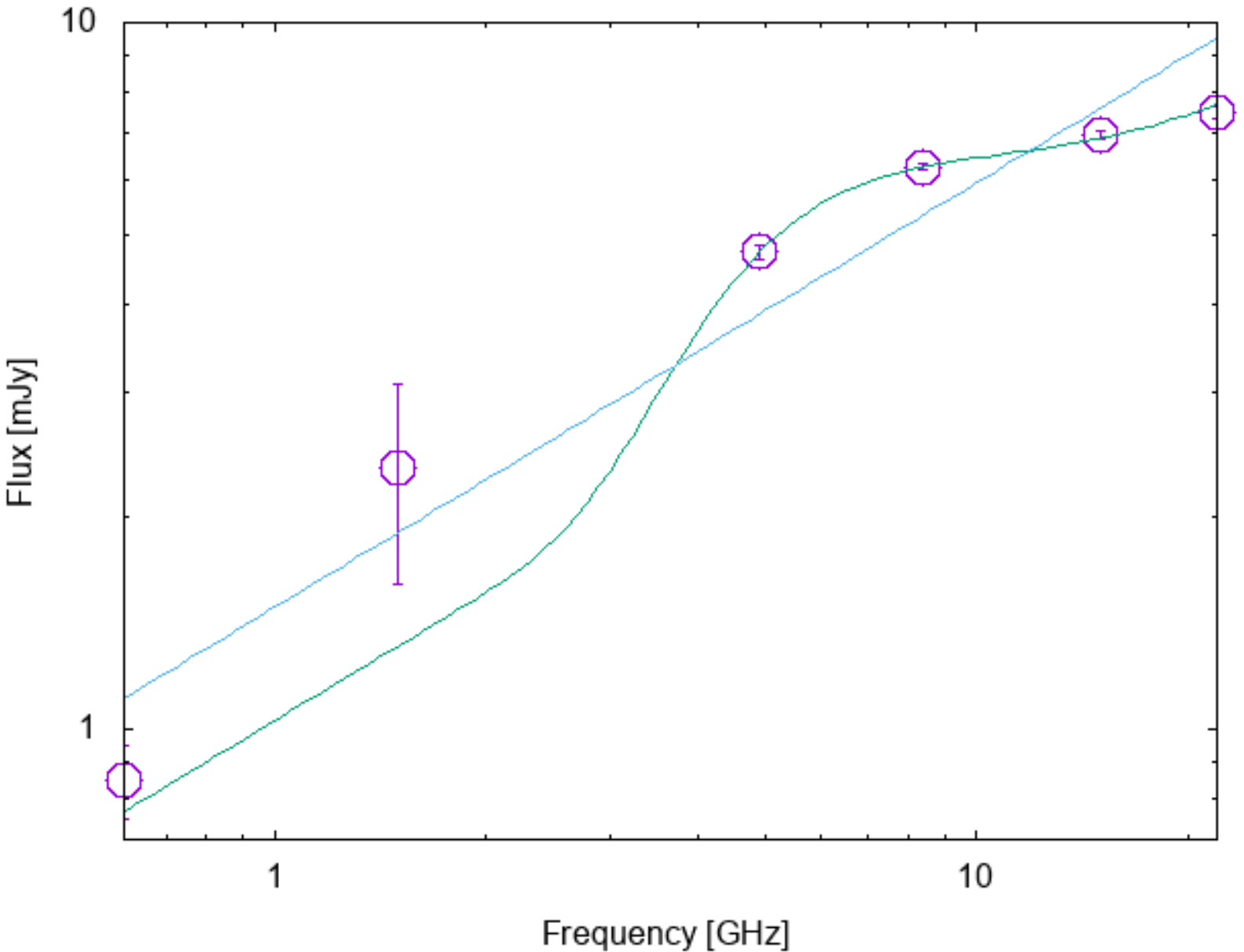}
\caption{Measured spectrum of WR\,140, at orbital phase 0.95 (magenta symbols). The cyan line represents thermal emission ($\alpha_{\rm ff}=0.6$). The green one, thermal emission  plus non-thermal emission with $\alpha_{\rm NT}=-0.6$ affected by FFA (see text).}
\label{fig:spectra140}
\end{center}
\end{figure}

For this system, we considered the { flux density} values given in \citet{Dougherty2005}. The authors observed the object with the Very Large Array at five bands spanning from 1.5 to 22~GHz. Table\,\ref{WRfluxes} lists the { flux densities} corresponding to phase 0.95, and the spectrum is shown in Fig.\,\ref{fig:spectra140}, top panel. We estimated the separation of the system components on the basis of the orbital solution \citep{Fahed2011}, using the most relevant inclination angle value available now. At the orbital phase of 0.95, the separation is about 1650 R$_\odot$. The position of the stagnation point, where ram pressures of both winds are balanced, is about 75\% of the stellar separation away from the WC star ($\sim$1237 R$_\odot$), and 25\% away from the O star ($\sim$413 R$_\odot$). For the calculations we assumed wind terminal velocities of 2860 and 3100~km~s$^{-1}$, and mass loss rate of 2~$\times$~10$^{-5}$ and 2~$\times$~10$^{-6}$~M$_\odot$~yr$^{-1}$, respectively for the WC and the O stars { \citep{Williams2011}}.

\begin{figure}[!h]
\begin{center}
\includegraphics[width=\columnwidth]{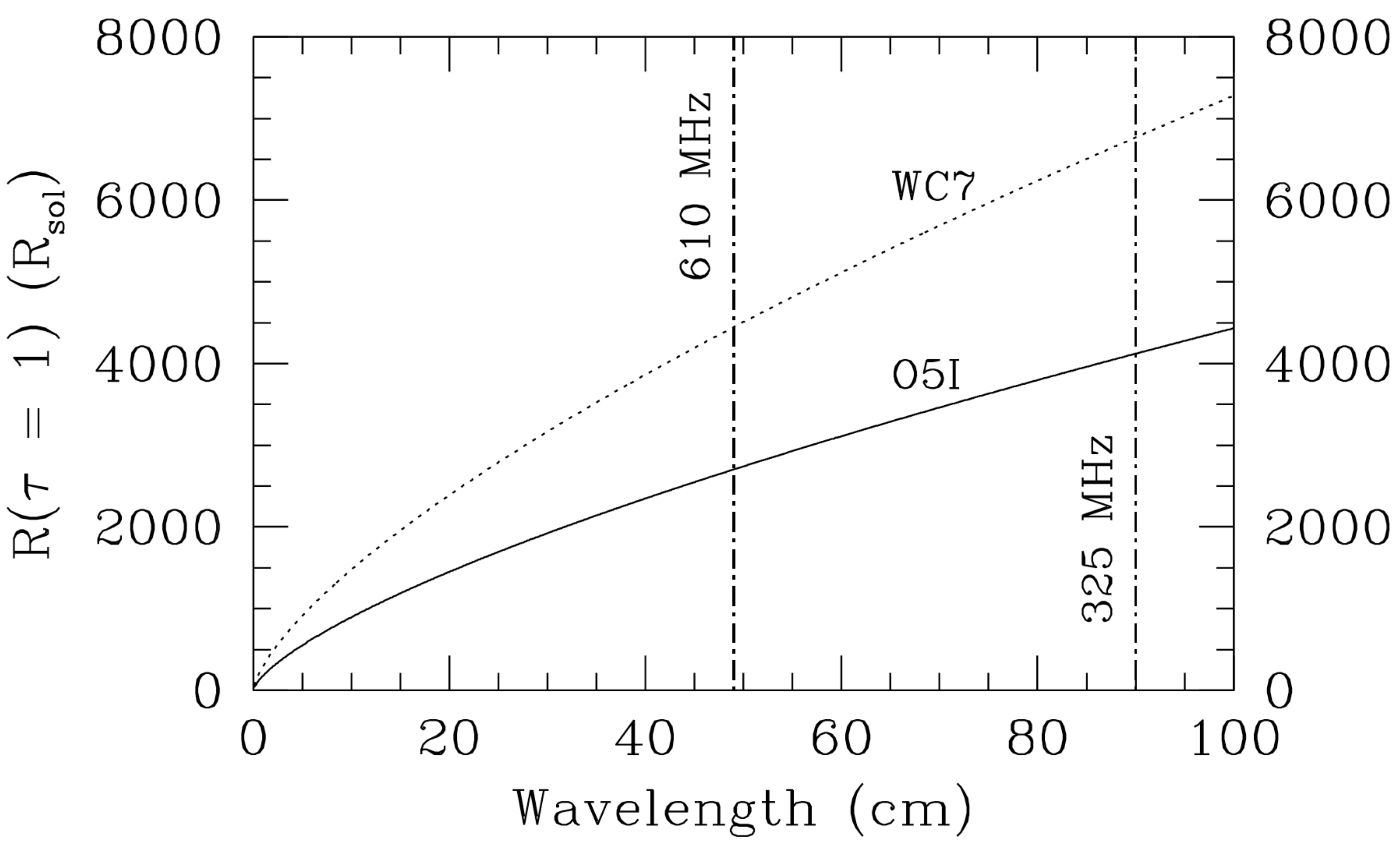}
\caption{Radio photosphere radius of the two components of WR\,140.
Vertical lines stand for our two observation wavelengths.}
\label{fig:radiophotos}
\end{center}
\end{figure}

Figure\,\ref{fig:radiophotos} is a plot of the estimated radial radio photosphere for a WC7 and a O5.5 star (assuming here a supergiant class), as a function of radio wavelength, based on the free-free emission theory developed by \citet{WB}. This approach has the benefit to provide a quick look at the likelihood that some synchrotron emission would emerge from the stellar winds \citep[see e.g.][for a recent use of this approach]{DeBecker2019b}. A more accurate quantification of the effect of free-free absorption would need a full radiative transfer treatment along the line of sight at a specific orbital phase, which is out of the scope of this paper. According to Figure\,\ref{fig:radiophotos}, the colliding-wind region will be inside the radio photospheres at 325, 610 and 1500~MHz frequencies of both stars, partially inside the O-star photosphere at 4.9 and 8.64 GHz, and outside the photosphere at 15 and 22 GHz. We will expect then that at lower frequencies, as the CWR is buried by the stellar photospheres, mostly the thermal emission from the winds (dominated by the WC contribution) can reach us; at medium frequencies, the spectrum has thermal contribution from the individual winds, but also non-thermal contribution from the partly unveiled CWR; and at higher frequencies, non-thermal emission from the CWR weakens to be disregarded, remaining the emission from the thermal winds which, besides, increases with frequency.

For WR\,140, the fit of Eq.\,\ref{eq:total-flux} with the mentioned routine failed to provide a function that represented the measured { flux densities}, but parameters with very large uncertainties, even fixing both the thermal and the non-thermal spectral indices. 
Figure\,\ref{fig:spectra140} shows the spectrum plotted together with only thermal emission, and with a combination of thermal plus non-thermal emission, this last affected by FFA. 

We note that deviations with respect to the simple model of Eq.\,\ref{eq:total-flux}
can be argued based on various facts. The actual scenario is much  more complex. The spectral index for spherical thermal winds can differ from its canonical value of 0.6 due to wind structure features like clumping, or ionization; more than one non-thermal emission region can be present, each one with its NT spectral index and normalization. Besides, the emitting region is extended, and the absorption for each volume element is not the same, but depends on the trajectory in the l.o.s. of the radiation, thus using a constant optical depth is an over simplification. Finally,  uncertainties in absolute calibrations of the flux densities measured with diverse instruments are expected to play a role.

\begin{table}[!h]
\caption{Radio {flux densities} of Wolf-Rayet systems.}\label{WRfluxes}
\centering
\begin{tabular}{l@{~~~}r@{~~~}c@{~~~}l}
\hline \hline
 Frequency  & $S_\nu$   & Synth.beam   & Observing \\
 $\nu$ (GHz)      & (mJy) & (arcsec$^2$)   & date(s)  \\
\hline
WR\,140$\dag$ &&& \\  
 \,\,\,\,\,0.61 & 0.9$\pm$0.1$^1$ & $6\times6$ & 17/07/2016\\ 
 \,\,\,\,\,1.5 & 2.35$\pm$0.7$^2$ & $46\times46$ & 27/09/2000 \\ 
 \,\,\,\,\,4.9 & 4.74$\pm$0.12$^2$ & $14\times14$ & 27/09/2000  \\ 
 \,\,\,\,\,8.4 & 6.24$\pm$0.06$^2$ & $8.2\times8.2$ & 27/09/2000  \\ 
 \,\,\,\,\,15 & 6.94$\pm$0.09$^2$ & $4.6\times4.6$ & 27/09/2000  \\ 
 \,\,\,\,\,22 & 7.48$\pm$0.14$^2$ & $3.1\times3.1$ & 27/09/2000 \\ 
 WR\,146 &&& \\
 \,\,\,\,\,0.15 & 35$\pm$10$^1$ & $25\times25$ & 2010--2012  \\ 
 \,\,\,\,\,0.325 & 111$\pm$2$^1$ & $10\times10$ & 06/02/2015 \\ 
 \,\,\,\,\,0.61 & 121$\pm$4$^1$ & $6\times6$ & 28/11/2014  \\
\,\,\,\,\,1.465 & 78.4$\pm$0.2$^3$ & $1.3\times1.3$ & 26/10/1996  \\
\,\,\,\,\,1.52 &  79.8$\pm$1.6$^4$ & $4.3\times4.3$  & 04/05/2015  \\ 
\,\,\,\,\,4.885 & 37.6$\pm$1.0$^3$ & $0.4\times0.4$ & 26/10/1996  \\
\,\,\,\,\,6.00 & 35.7$\pm$0.7$^4$ & $1\times1$ & 04/05/2015 \\
\,\,\,\,\,8.435 & 29.8$\pm$0.8$^3$ & $0.24\times0.24$ & 26/10/1996 \\
\,\,\,\,\,22.46 & 17.4$\pm$1.5$^3$ & $0.1\times0.1$ & 26/10/1996 \\
WR\,147 &&& \\ 
\,\,\,\,\, 0.325 & 16$\pm$4$^5$ & $55\times55$  & 1995  \\ 
 \,\,\,\,\, 0.61 & 20.6$\pm$0.3$^1$ &  $6\times6$ & 28/11/2014 \\ 
 \,\,\,\,\,1.4 & 25.5$\pm0.5$$^{6,7}$ &  $12.5 \times 16.3$ & 1989--1997 \\ 
 \,\,\,\,\,4.86 & 35.4$\pm$0.4$^7$ &  $3.5 \times 5$ & 1988--1997  \\
 \,\,\,\,\,8.44 & 40.3$\pm$4$^6$ & $1.1 \times 0.8$ & 02/11/1995  \\ 
 \,\,\,\,\,14.94 & 46.2$\pm$3$^6$ & $0.7 \times 0.5$ & 02/11/1995 \\
 \,\,\,\,\,22.46 & 52$\pm$5$^6$ & $0.4 \times 0.3$ & 02/11/1995 \\
 \,\,\,\,\,42.8 & 82.8$\pm$2$^8$ & 0.3--1 & 1994--1995 \\  
\hline \hline
\end{tabular}
\tabnote{$\dag$: In the case of WR\,140, the {flux densities} correspond to phase 0.95. References: 1 = this work; 2 = \citet{Dougherty2005}; 3 = \citet{Dougherty2000}; 4 = \citet{Hales2016}; 5: \citet{Setia2003};  6 = \citet{Skinner1999}: 7 = \citet{Setia2001}; 8 = \citet{Contreras1996}. In the case of data from 6, we quote the { flux densities} derived with IMFIT, but with an error interval to include the TVSTAT values.
}
\end{table}

\begin{table}[!h]
\caption{Radio { flux densities} of OB+OB systems.}\label{OBfluxes}
\centering
\begin{tabular}{l@{~~~}r@{~~~}c@{~~~}l}
\hline \hline
 Frequency  & $S_\nu$  & Synth.beam   & Observing \\
$\nu$ (GHz)      & (mJy) & (arcsec$^2$)   & date(s)  \\
\hline
OB2-335 &&& \\ 
\,\,\,\,\, 0.325 & 2.6$\pm$0.3$^1$ & $10\times10$  & 04/11/2013  \\ 
\,\,\,\,\, 0.61 & 5.4$\pm$0.2$^1$ &  $6\times6$ & 29/11/2014 \\ 
\,\,\,\,\, 1.4 & 3.6$\pm$0.7$^2$ & $13\times13$ & 1996-1997 \\
\,\,\,\,\, 4.886 & 1.3$\pm$0.1$^2$ & not prov. & 25/02/2001 \\
\,\,\,\,\,  8.44 & 1.6$\pm$0.3$^2$ & not prov. & 25/02/2001 \\
MWC\,349 &&&\\
\,\,\,\,\, 0.325 & 28.6$\pm$1.21$^1$ & $10\times10$ & 04/11/2013  \\ 
\,\,\,\,\, 0.61 & 40.5$\pm$0.54$^1$ & $6\times6$ & 29/11/2014 \\ 
\,\,\,\,\, 1.425 & 76.4$\pm$6.4$^3$ & $1.3\times1.2$ & 05/12/1996 \\
\,\,\,\,\, 2.695 & 100$\pm$15$^4$ & $3 \times 3$ & 1972-1973\\
\,\,\,\,\, 4.85 & 154.8$\pm$8.5$^3$ & $0.4\times0.4$ & 05/12/1996 \\
\,\,\,\,\, 8.31 & 183.5$\pm$9.5$^3$ & $0.2\times0.2$ & 31/12/1988 \\
\,\,\,\,\, 14.94 & 380.0$\pm$21.2$^3$ & $0.1\times0.1$ & 05/12/1996 \\ 
\,\,\,\,\, 22.367 & 446.2$\pm$44.8$^3$ & $0.09\times0.08$ & 29/03/1990 \\
\,\,\,\,\, 43.34 & 635.0$\pm$95.6$^3$ & $0.04\times0.03$& 16/12/1996 \\
\hline \hline
\end{tabular}
\tabnote{not prov.: not provided by authors. References: 1 = this work; 2 = \citet{Setia2003}; 3: \citet{tafoya2004}; 4 = \citet{hjellming1973}.}
\end{table}

\subsubsection{WR\,146} 

To study the radio spectrum of this system we included our measurements at the GMRT bands of 150, 325 and 610 MHz and the { flux densities} measured by \citet{Dougherty2000} and by \citet{Hales2016} using the Very Large Array between 1.5 and 22.5 GHz. 
The 168-mas separation between resolved components of WR\,146 measured by \citet{Niemela1998} 
and the 116-mas separation between the southern thermal source (the WR thermal wind) and the northern non-thermal source (the colliding-wind region) reported from MERLIN 5-GHz observations \citep{Dougherty1996} allowed to derive the distances of the WR and O-type stars to the CWR, $D_{\rm WR-CWR}$ and $D_{\rm O-CWR}$. 

We detected the system down to 150\,MHz ($\lambda$=2\,m). For a WC6 star, the radio photosphere at this frequency is about 10\,000~R$_\odot$, if we adopt a terminal velocity of 2200 km~s$^{-1}$ and a mass loss rate of 1.3~$\times$~10$^{-5}$ M$_\odot$~yr$^{-1}$. This value is much lower than $D_{\rm WR-CWR}$ for a system distance of 1.1~kpc (27\,500~R$_\odot$). Regarding the primary star, previous works show important  uncertainty on the nature of the so-called ``O-type companion'', which may even be a binary. In such circumstances, it is not possible to make any valid estimate of the photosphere. However, if one stays in the O-type regime
with no hidden WR component, one can clearly state that the extension of the photosphere will be significantly smaller than that of the WC star. For this particular system, some spectral features seem to favor a later type object, which would decrease further its size (especially because of a lower mass loss rate). Then, as $D_{\rm O-CWR}$ = $\sim$13\,750~R$_\odot$,  
the CWR will not be buried in the stellar photospheres, even at the lowest frequency (largest photosphere).

\begin{figure}[!t]
\begin{center}
\includegraphics[width=8cm]{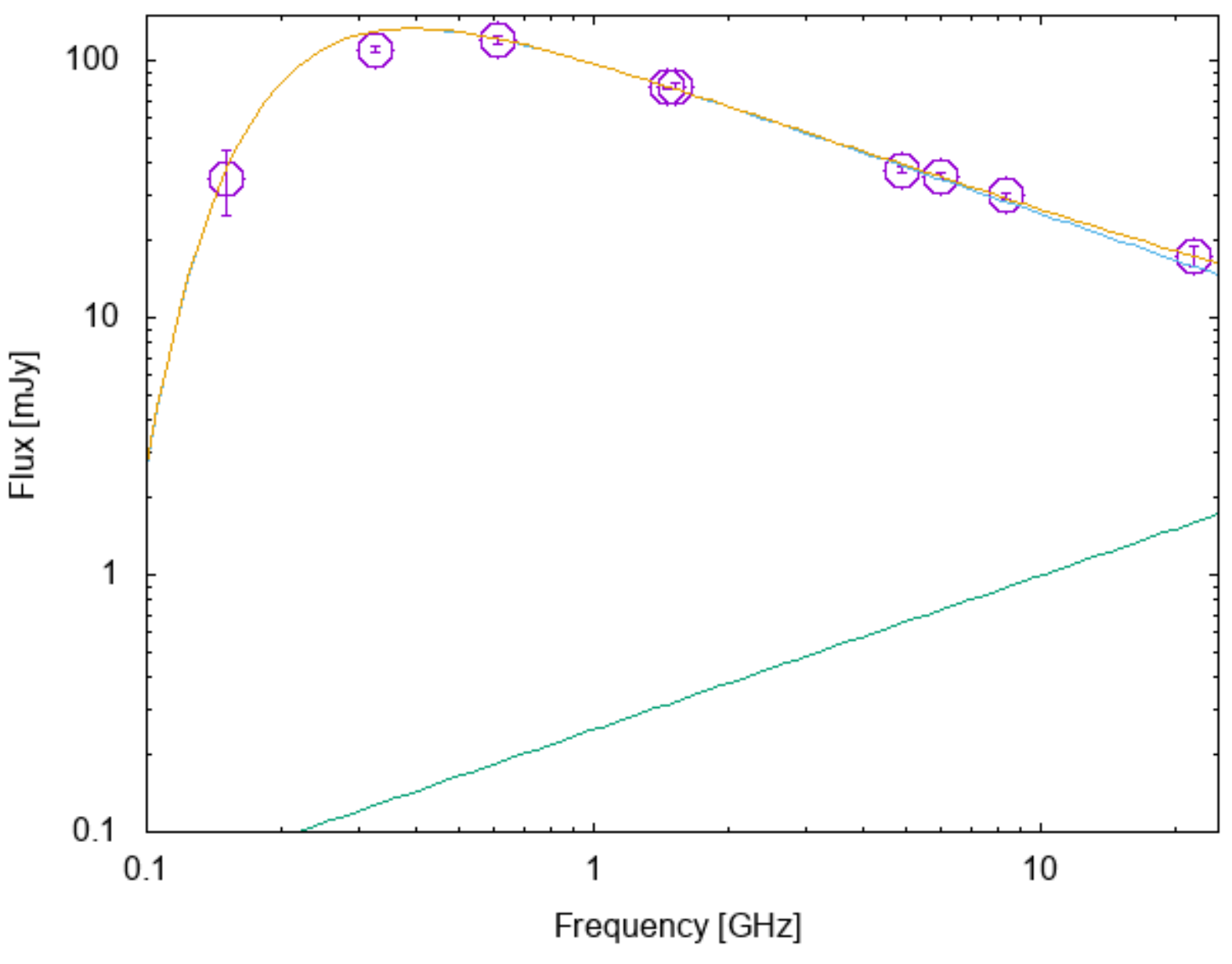}
\caption{Measured spectrum of WR\,146 (magenta symbols): 
free-free absorption up to 610~MHz (cyan), thermal emission from the WC6 wind (green, $\alpha=0.6$) and non-thermal emission from the CWR with $\alpha=-0.6$  \citep[orange, see][]{Hales2016}.}
\label{fig:spectra146}
\end{center}
\end{figure}

The spectrum of WR\,146 presented in Fig.\,\ref{fig:spectra146} shows non-thermal emission, very strong at GHz frequencies. Two different trends can be appreciated: below and above $\nu=610$~MHz. PACWBs with exposed colliding-wind regions are expected to radiate synchrotron emission, with a spectral index $\alpha_{\rm NT}$ related to the electron  index $p$ of the relativistic electron (power-law) distribution, such as $\alpha_{\rm NT} = -(p - 1)/2$. For a strong adiabatic shock in a CWR, Fermi acceleration characterized with $p=2$ is expected \citep{Bell1978a}, thus $\alpha_{\rm NT}=-0.5$. \citet{Hales2016} observed WR~146 at 1.4 and 6~GHz every 64-MHz sub-bands, and found best fits for $\alpha_{\rm NT} \approx -0.6$. 
We then adopt for this system, and above 610~MHz, non-thermal emission originated at the CWR $\propto \nu^{-0.6}$. Added to that, we adopt a thermal contribution from the individual winds of the resolved components, considering that it represents 1~mJy at 5~GHz as measured by \citet{Dougherty1996} with $\alpha=0.6$ (see Fig.\,\ref{fig:spectra146}). The fit of the spectrum fixing the spectral indices $\alpha_{\rm ff}$ and $\alpha_{\rm NT}$ of Eq.\,\ref{eq:total-flux} provided the parameters { A$ = 0.4\pm1.2$~mJy~GHz$^{-1}$, B$ = 100\pm2$~mJy~GHz$^{-1}$} and $\tau_0 = 0.05\pm0.01$, see Fig\,\ref{fig:spectra146}.

\citet{Setia2000} performed a thorough study of this system, from its structure to its energetics. By applying Eq.\,\ref{eq:freqSSA} they conclude that SSA is not relevant, and that the turnover is most probably due to FFA. The results of the present fit to the spectrum, that includes the measurement at 150~MHz,  explains the turnover in terms of FFA, reinforcing their findings. If this is the case, the results also imply that the RTe, if relevant, will be acting below 150~MHz. The electron density at the CWR derived by \citet{Setia2000}, scaled to a distance of 1.1~kpc, is $\sim$15\,100~cm$^{-3}$. Following Eq.\,\ref{eq:freqRTe}, we obtained the lower limit for the magnetic field strength, as $M > 2$~mGauss.

\subsubsection{WR\,147}

\begin{figure}[!t]
\begin{center}
\includegraphics[width=8cm]{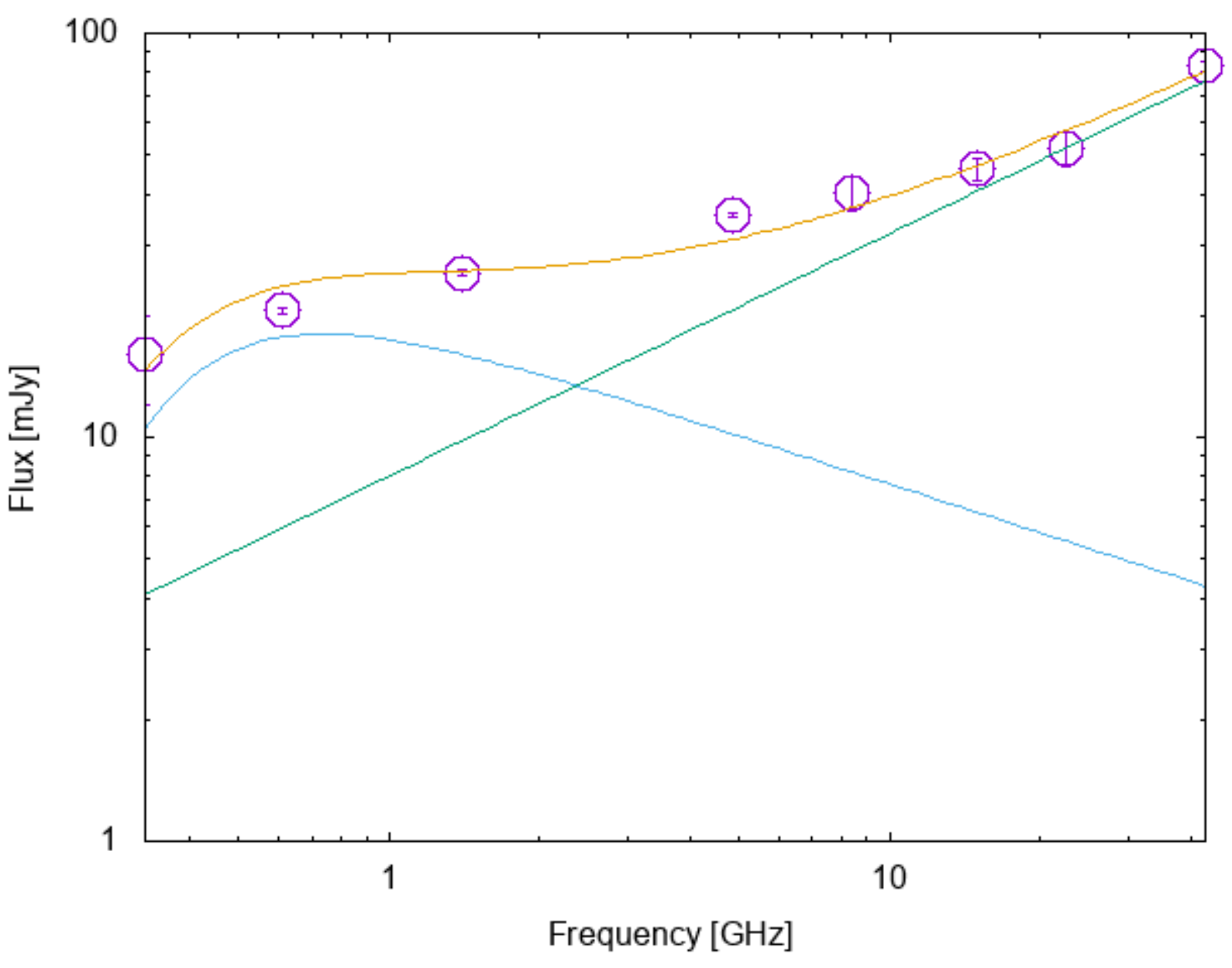}
\caption{Spectrum of WR\,147 (magenta symbols); in orange, the best fit adding thermal emission as 8.8\,$\nu^{0.6}$ (green) plus non-thermal emission affected by FFA as  21\,$\nu^{-0.4}\,\exp(-0.2\nu^{-2.1})$ (cyan).}
\label{fig:spectra147}
\end{center}
\end{figure}

To analyze the radio spectrum of the WR\,147 system, we considered the 610-MHz { flux density} value reported here, together with those at other bands measured by \citet{Setia2001}, \citet{Skinner1999}, and \citet{Contreras1996}. Figure \,\ref{fig:spectra147} demonstrates that the GMRT { flux density} value 
perfectly follows the curves already obtained by the papers cited above,
confirming their findings (see Fig.\,\ref{fig:spectra147}). We fit a two-term spectrum, with a thermal contribution of $\alpha=0.6$, and a non-thermal one modified by FFA. The parameters of the  fit resulted as 
{ A$ = 8.8\pm1.0$~mJy~GHz$^{-1}$, B$ = 21\pm4$~mJy~GHz$^{-1}$}, $\alpha_{\rm NT} = -0.4\pm0.25$,  $\tau_0 = 0.2\pm0.1$, and the curves are shown in Fig.\,\ref{fig:spectra147}.

\subsubsection{Cyg\,OB2-335}

\begin{figure}[!t]
\begin{center}
\includegraphics[width=8cm]{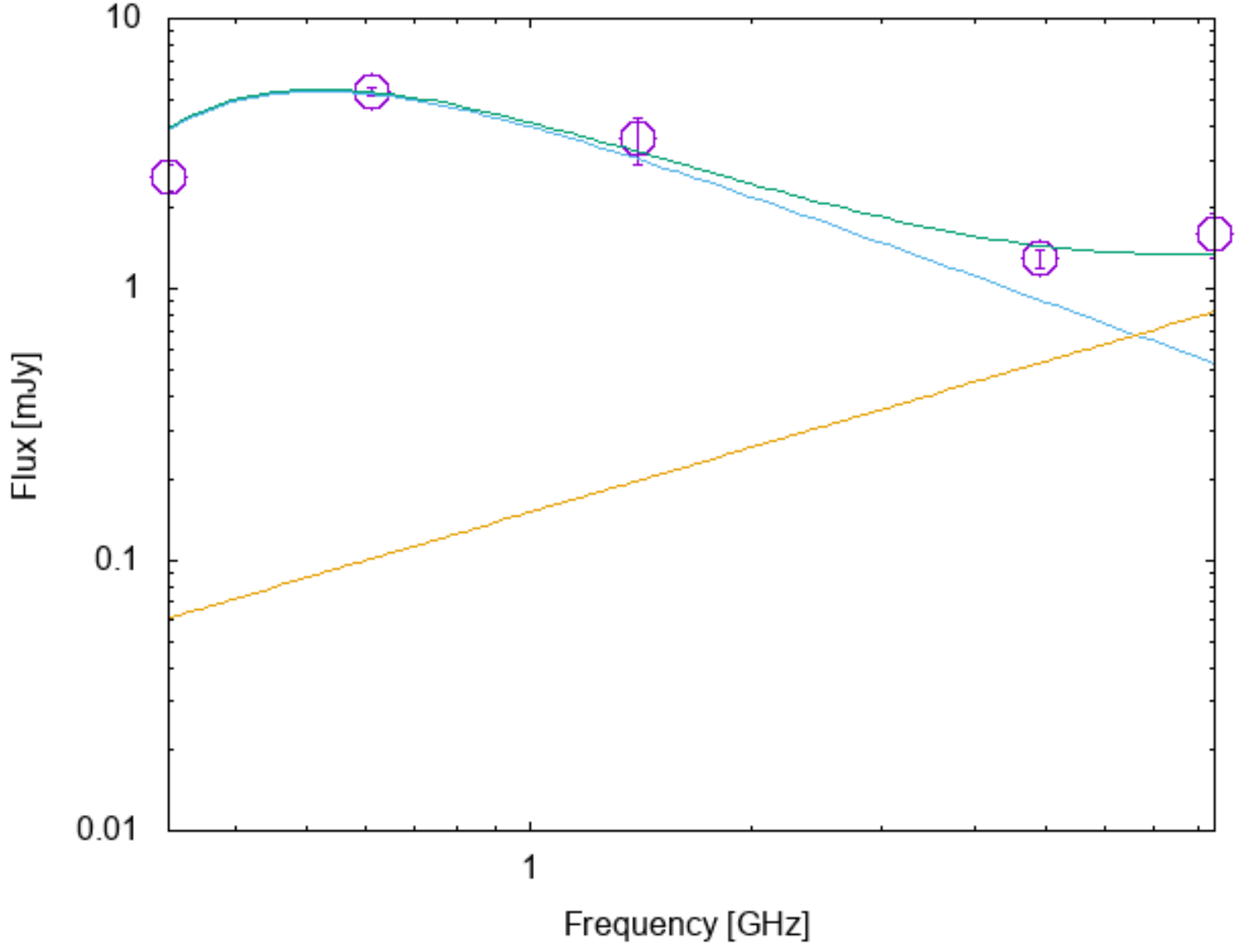}
\caption{Measured spectrum of Cyg\,OB2-335 (magenta symbols); in green, the best fit adding thermal emission as 0.14\,$\nu^{0.7}$ (orange) plus non-thermal emission affected by FFA as  4.5\,$\nu^{-0.95}\,\exp(-0.13\nu^{-2.1})$ (cyan).}
\label{fig:spectra335}
\end{center}
\end{figure}

Figure\,\ref{fig:spectra335} depicts the radio { flux densities} measured by \citet{Setia2003} and this work. The system has a period of 2.8~d \citep{Kobulnicky2014}, which makes very difficult to reconcile the negative spectral index quoted between measurements at 1.4 and 4.9~GHz by \citet{Setia2003}, since synchrotron radiation could hardly escape from being absorbed by the optically  thick thermal winds of the binary  components. \cite{DeBecker2017} proposed to investigate the object in searching for a potential third component of the system in a wider orbit, capable of developing an unveiled CWR. In such a case, and if the third orbit is large enough for the measurements given in Table\,\ref{WRfluxes} (along 1996 to 2014) to be comparable, then we can fit the spectrum using Eq.\,\ref{eq:total-flux}. Due to the low number of points, and the large errors for some of them, the fitting routine provided parameters without the covariance matrix. The resulting expression was $S_\nu = 0.14\,\nu^{0.7}+ 4.5\,\nu^{-0.95}\,\exp(-0.13\nu^{-2.1})$. Figure\,\ref{fig:spectra335} 
shows the problems in fitting the { flux density} at 325~MHz, taking into account just FFA. { Alternatively, let us explore the idea that the turnover is due to SSA. We do not have any direct information on the size of the emitting region ($\theta$ in Eq. 2). One can just say that, as some synchrotron radiation is detected, the source should be significantly located out of the radio photosphere. This does not tell us about its exact size, but its extension may be of the same order. We thus determined the size of the radio photosphere of the O7\,V component of the system. For typical mass loss rate and terminal velocity values of 2.4\,$\times$\,10$^{-7}$\,M$_\odot$\,yr$^{-1}$ and 2500\,km\,s$^{-1}$ \citep{Vink2000} we get $\theta \sim 1000$ R$_\odot$. In the framework of these assumptions, if SSA is in action, the magnetic field strength is of the order of 1~mGauss. If the assumed size is 5 times smaller, the magnetic field strength is 3~$\mu$Gauss. Even though such values are compatible with expectations from a CWR, we caution that their determination relies on strong assumptions. In particular, there is not warranty that SSA is active and the emission region size is still undetermined.}

\subsubsection{MWC\,349}
\label{spectrum349}

The B[e] star MCW\,349A has been largely studied by means of radio observations at many bands, at first due to its high brightness. \citet{cohen1985} discovered a bipolar emitting region or wind at 15~MHz of ``hourglass'' shape, and a kind of bridge at 5~GHz that separates it from its proposed companion, MWC\,349B. \citet{tafoya2004} presented observations taken with the Very Large Array at 330~MHz up to 44.34~GHz, from which  they measured the corresponding { flux densities}. Using data from 1.425 to 44.34~GHz the authors calculated the spectral index of the ionized wind as $0.64\pm0.03$, and proposed that the ionized emission came from the photoevaporation of a disk in the equatorial plane of the star. According to the investigation of  \citet{luisfelipe2007}, the former shape of the bipolar outflow has  been changing to an almost square shape.

\begin{figure}[!t]
\begin{center}
\includegraphics[width=8cm]{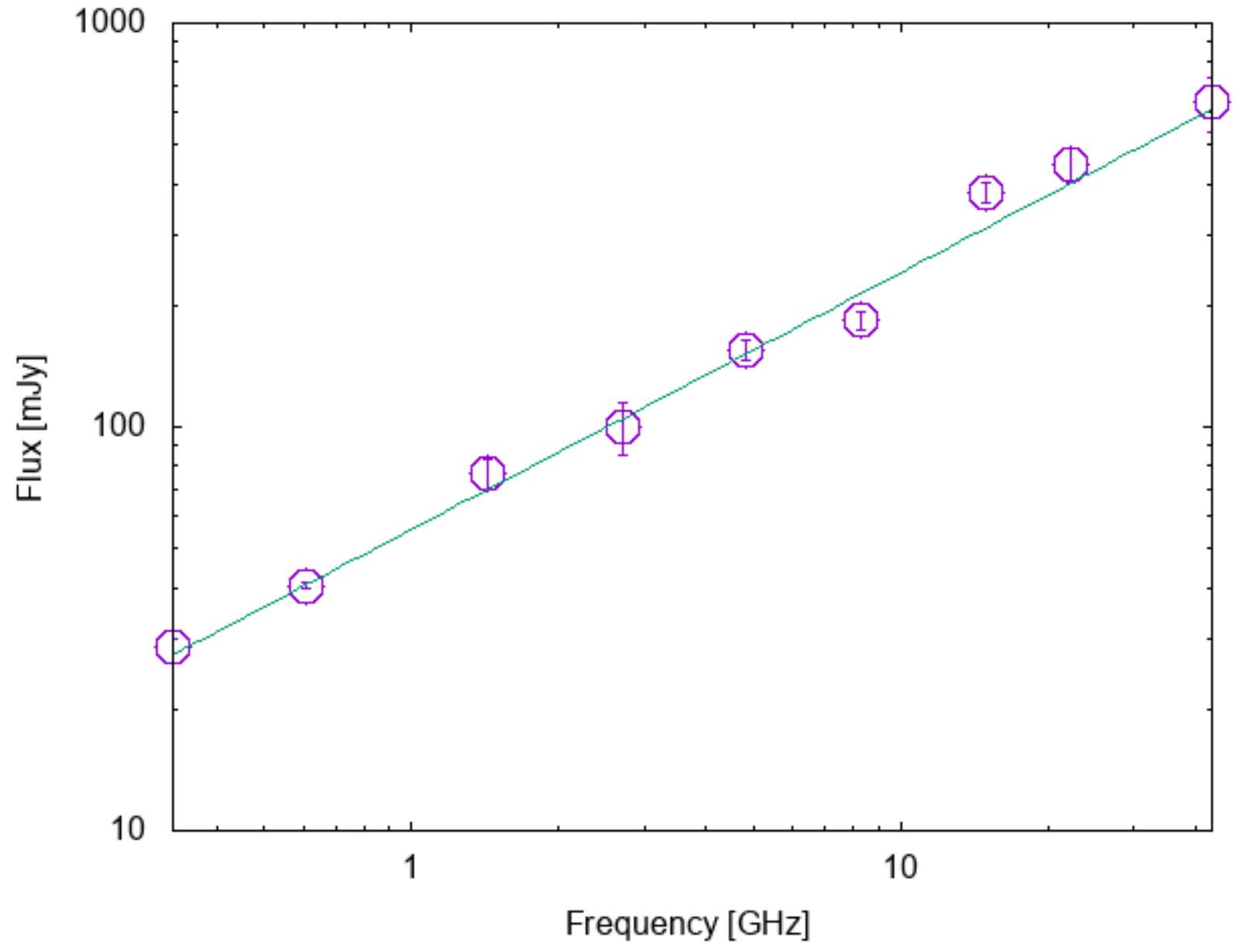}
\caption{Measured spectrum of MWC\,349 (magenta symbols); in green, the best fit considering thermal emission as 55.7\,$\nu^{0.64}$.}
\label{fig:spectra349}
\end{center}
\end{figure}

We complemented the {flux densities} from 1.4 to 44~GHz given by \citet{tafoya2004} with the ones obtained here, and a 2.7~GHz {flux density} value given by \citet{hjellming1973} (see Table\,\ref{OBfluxes}). The set allowed us to fit a thermal contribution  to  the spectrum in the form of $S_\nu = {\rm A}\,\nu^{\alpha_{\rm ff}}$, with { A$ = 55.71\pm1.21$~mJy~GHz$^{-1}$} and $\alpha_{\rm ff} = 0.636\pm0.021$, confirming the previous result. In Fig.\,\ref{fig:spectra349} we displayed the observed { flux densities} and the fit function.  

\section{CONCLUSIONS}\label{concl}

We reported on the investigation of a 15-square degrees survey of the Cygnus region using the GMRT, at 325\,MHz and 610\,MHz frequencies. Our main objective was to investigate the low frequency radio emission from the population of massive stars in that region, with emphasis on O-type and Wolf-Rayet stars, along with a couple of B-type remarkable objects.

\begin{enumerate}
\item[1.] Our investigation allowed us to provide evidence that massive stars, and in particular colliding-wind binary systems, are able to emit radio continuum down to 150~MHz. Out of nine WR objects in the field, we detected four of them at least at one frequency. For O-type stars, we detected seven objects out of 87 present in the field of our survey. We interpret the low detection rate for O-type stars in terms of (i) thermal emission much fainter at low frequencies, (ii) strong free-free absorption by the stellar winds that is strongly attenuating any putative synchrotron emission (for binary systems), and (iii) low non-thermal energy budget in a population dominated by later type objects characterized by a lower wind kinetic power.

\item[2.] The measurements at these frequencies appear also to be especially appropriate to investigate synchrotron emission from colliding-wind binaries. Our data enabled us to identify two new members in the class of PACWBs, namely ALS\,15108\,AB and Cyg\,OB2-12. With these two additional members, the total number of known PACWBs amounts to about 45 objects so far.

\item[3.] We also focused on the Wolf-Rayet systems WR\,140, WR\,146, and WR\,147 to discuss their main spectral properties, making use of broad band radio measurements including our new measurements and previous ones obtained at higher frequencies. In general, the spectral shape can be well interpreted in term of free-free absorbed synchrotron emission at lower frequencies, on top of a thermal emission from the stellar winds. In this context, the extension of the radio photosphere and the geometry of the system play a key role. WR\,146, the only system with {flux density} measurements down to 150~MHz, presents a clear turnover. These results may indicate that the Razin-Tsytovitch effect does not seem to be relevant at the frequencies investigated.

\item[4.] For the O$+$O system Cyg\,OB2-335, other processes than FFA have to be invoked to match the observations, but the current information on this system precludes further analysis. Dedicated, simultaneous  observations, at the highest angular resolution and sensitivity are crucial to disentangle the processes at work in this system, and search for the evidence of a third companion.

\item[5.] Finally, the high signal-to-noise {flux density} measurements at MHz frequencies provided here for the B$+$B system MWC\,349 confirm the thermal nature of its emitting wind.

\end{enumerate}

\begin{acknowledgements}
The authors are grateful to the referee, whose comments and suggestions resulted in improving the analysis and presentation  of the article. The radio data presented here were obtained with the Giant Metrewave Radio Telescope (GMRT). The GMRT is operated by the National Centre for Radio Astrophysics of the Tata Institute of Fundamental Research. We thank the staff of the GMRT that made these observations possible. P.B. acknowledges support from ANPCyT PICT 0773--2017, ME. Colazo for computing assistance,  S. del Palacio for fruitful discussions and the staff at NCRA, Pune, for a wonderful stay. This research has made use of the SIMBAD database, operated at CDS, Strasbourg, France, and of NASA's Astrophysics Data System bibliographic services. 
\end{acknowledgements}

\bibliographystyle{pasa-mnras}
\bibliography{biblio}


\begin{appendix}

\section{NON-DETECTED O-TYPE STARS}

Here we present Table\,\ref{nondet}, with the undetected O stars: star name, spectral type classification and its reference, and coordinates. The rms was estimated using the image at 610~MHz, over an area of $\sim 4' \times 4'$, computed over four boxes around the star. 

\begin{table*}
\caption{Non detected O-type stars in the area of the Cygnus region observed in the present study.} 
\centering
\begin{tabular*}{\textwidth}{@{}l\x l\x l\x c\x r@{}}
\hline \hline
Name &  Spectral type & Reference & $RA, Dec$ (J2000) & rms at 610~MHz \\
     &   classification &         &[h,m,s],[d,','']  & (mJy~beam$^{-1}$) \\
\hline%
HD\,193117 & O9.5II & (1) & 20:16:59.89,$+$40:50:37.4 & 0.10 \\
LS\,III$+$41 14 & O9.5V(n) & (2) & 20:17:05.52,$+$41:57:46.9 & 0.10 \\
HD\,228759 & O6.5V(n)((f))z & (2) & 20:17:07.54,$+$41:57:26.5 & 0.10 \\
HD\,193322AaAb & O9IV(n)+B1.5V & (2) & 20:18:06.99,$+$40:43:55.5 & 0.15  \\
TYC\,3159--6--1 & O9.5-O9.7Ib & (3) & 20:18:40.34,$+$41:32:45.1 & 0.10 \\ 
J20190610$+$4037004 & O9.7Iab & (4) & 20:19:06.10,$+$40:37:00.4 & 0.30 \\ 
J20194916$+$4052090 & O9.5V & (4) & 20:19:49.15,$+$40:52:08.9  & 0.25 \\ 
ALS\,11244 & O4.5III(n)(fc)p & (2) & 20:22:37.78,$+$41:40:29.2 & 0.10 \\
HD\,229196 & O6II(f) & (2) & 20:23:10.79,$+$40:52:29.9 & 0.15 \\
ALS\,19302 & O9.5:V & (5) & 20:23:14.55,$+$40:45:19.1 & 0.20 \\
HD\,229202 & O7.5V(n)((f)) & (2) & 20:23:22.84,$+$40:09:22.5 & 0.80 \\
ALS\,11321 & O9III & (4) & 20:25:06.52,$+$40:35:49.8 & 0.20 \\
J20272428$+$4115458 & O9.5V & (4) & 20:27:24.28,$+$41:15:45.8 & 0.10 \\
BD$+$40\,4179 & O8Vz & (2) & 20:27:43.62,$+$40:35:43.5 & 0.15 \\
J20283039$+$4105290 & OC9.7Ia & (4) & 20:28:30.39,$+$41:05:29.1 & 0.12 \\
ALS\,11376 & O7 & (4) & 20:28:32.03,$+$40:49:02.9 & 0.40 \\
ALS\,11378 & O8.5V & (4) & 20:28:40.81,$+$43:08:58.5 & 0.10 \\
J20293480$+$4120089 & O9.5V & (4) & 20:29:34.80,$+$41:20:08.9 & 0.15 \\ 
CPR2002-A18 & O8V & (4) & 20:30:07.88,$+$41:23:50.4 & 0.10 \\
J20301839$+$4053466 & O9V & (4) & 20:30:18.39,$+$40:53:46.6 & 0.20 \\ 
Cyg\,OB2-B17 & O6Iaf$+$O9:Ia: & (2) & 20:30:27.30,$+$41:13:25.3 & 0.15 \\
HD\,195592 & O9.7Ia & (6) & 20:30:34.97,$+$44:18:54.9 & 0.15 \\
ALS\,15129 & O6.5V & (4) & 20:30:39.80,$+$41:36:50.7 & 0.10 \\
CPR2002-A26 & O9V & (4) & 20:30:57.72,$+$41:09:57.5 & 0.10 \\
ALS\,19637 & O7V((f)) & (4) & 20:31:00.20,$+$40:49:49.7 & 0.10 \\
Cyg\,OB2-1 & O8IV(n((f)) & (2) & 20:31:10.54,$+$41:31:53.5 & 0.10 \\
ALS\,15133 & O9.5IV & (2) & 20:31:18.33,$+$41:21;21.7 & 0.10 \\
ALS\,21081 & O7Ib(f) & (4) & 20:31:36.91,$+$40:59:09.0 & 0.10 \\
Cyg\,OB2-3A & O8.5Ib(f)$+$O6III: & (2) & 20:31:37.51,$+$41:13:21.0 & 0.15 \\
ALS\,15116 & O8V & (4) & 20:31:45.40,$+$41:18:26.7 & 0.10 \\
Cyg OB2-20 & O9.7IV & (2) & 20:31:49.67,$+$41 28:26.5 & 0.10 \\
GOS\,G080.03$+$00.94\,01 & O7.5Vz & (2) & 20:31:59.61,$+$41:14:50.5 & 0.10  \\ 
Cyg\,OB2-4A & O7III((f)) & (2) & 20:32:13.83,$+$41:27:12.0 & 0.10 \\
Cyg\,OB2-14 & O9V & (7) & 20:32:16.56,$+$41:25:35.7 & 0.10 \\
RLP\,145 & O9.5V & (8) & 20:32:19.75,$+$41:44:46.9 & 0.15 \\ 
Cyg\,OB2-15 & O8III & (2)  & 20:32:27.67,$+$41:26 22.1 & 0.10 \\
ALS\,19628 & O9.5IV & (4) &  20:32:30.31,$+$40:34:33.2 & 0.10 \\
CPR2002-A38  & O8V & (2) & 20:32:34.87,$+$40:56:17.3  & 0.12 \\
CPR2002-A25 & O8III & (2) & 20:32:38.44,$+$40:40:44.5 & 0.12 \\
Cyg\,OB2-16 & OO7.5IIV(n) & (2) & 20:32:38.56,$+$41:25:13.8 & 0.11 \\
Cyg\,OB2-6&O8.5V(n) & (2) & 20:32:45.45,$+$41:25:37.5 & 0.10 \\
Cyg\,OB2-17 & O8V & (2) & 20:32:50.01,$+$41:23:44.7 & 0.10 \\
ALS\,15111 & O8V & (2) & 20:32:59.19,$+$41:24:25.5 & 0.10 \\
Cyg\,OB2-41 & O9.7III(n) &  (2) &  20:32:59.64,$+$41:15:14.7 & 0.12 \\
ALS\,15131 & O7.5V((f)) & (2) & 20:33:02.92,$+$41:17:43.1 & 0.13 \\
Cyg\,OB2-22A & O3If* & (2) & 20:33:08.76,$+$41:13:18.6 & 0.10 \\
Cyg\,OB2-22B & O6V((f)) & (2) & 20:33:08.84,$+$41:13:17.4 &  0.10 \\
Cyg\,OB2-22C &  O9.5IIIn & (2) & 20:33:09.60,$+$41:13:00.5 & 0.10 \\
Cyg\,OB2-50 & O9.5IIIn & (6) & 20:33:09.60,$+$41:13:00.6 & 0.10 \\
\hline \hline
\end{tabular*}\label{nondet}\end{table*}

\setcounter{table}{4}

\begin{table*}
\caption{(continued).} 
\centering
\begin{tabular*}{\textwidth}{@{}l\x l\x l\x c\x r@{}}
\hline \hline
Name &  Spectral type & Reference & $RA, De$ (J2000) & rms at 610~MHz \\
     &   classification &                &[h,m,s],[d,','']  & (mJy~beam$^{-1}$)\\
\hline%
Cyg OB2-22D & O9.5Vn & (2) & 20:33:10.12,$+$41:13:10.1 & 0.10 \\
Cyg OB2-9 & O4.5If & (2) & 20:33:10.73,$+$41:15:08.2 & 0.12 \\
ALS 15148 & O6.5:V & (2) & 20:33:13.27,$+$41:13:28.7 & 0.10 \\
ALS 15128 & O8V & (4) & 20:33:13.69,$+$41:13:05.8 & 0.10 \\
Cyg OB2-7 & O3If* & (2) & 20:33:14.12,$+$41:20:21.9 & 0.11 \\
Cyg OB2-8B & O6II(f) & (2) & 20:33:14.76,$+$41:18:41.8 & 0.11 \\
Cyg OB2-23 & O9V & (7) & 20:33:15.71,$+$41:20:17.2 & 0.11 \\
Cyg OB2-8D & O8.5V(n) & (2) & 20:33:16.33,$+$41:19:02.0 & 0.11 \\
Cyg OB2-24 & O8V(n) & (2) & 20:33:17.48,$+$41:17:09.3 & 0.14 \\
Cyg OB2-8C & O4.5(fc)p var & (2) & 20:33:17.98,$+$41:18:31.2 & 0.14 \\
ALS 15115 & O8V & (2) & 20:33:18.05,$+$41:21:36.9 & 0.12 \\
ALS 15123 & O9V & (4) & 20:33:21.02,$+$41:17:40.1 & 0.14 \\
Cyg OB2-25 A & O8.5Vz & (2) & 20:33:25.54,$+$41:33:27.0 & 0.10  \\
ALS 15134 & O8.5Vz & (2) & 20:33:26.75,$+$41:10:59.5 & 0.10 \\
Cyg OB2-74 & O8V & (4) & 20:33:30.32,$+$41:35:57.9 & 0.10 \\
Cyg OB2-70 & O9.5IV(n) & (2) & 20:33:37.00,$+$41:16:11.3 & 0.12 \\
ALS 15124 & O7V & (4) & 20:33:40.86,$+$41:30:18.9 & 0.10 \\
Cyg OB2-10 & O9.7Iab & (2) & 20:33:46.11,$+$41:33:01.1 & 0.11 \\
J20335842$+$4019411 & O9: & (4) & 20:33:58.42,$+$40:19:41.2 & 0.14 \\
Cyg OB2-27AB & O9.7V(n)$+$O9.7V:(n) & (2) & 20:33:59.53,$+$41:17:35.5 & 0.12 \\
ALS 15145 & O8.5V & (4) & 20:34:04.86,$+$41:05:12.9 & 0.14 \\
Cyg OB2-11 & O5.5Ifc & (2) &  20:34:08.51,$+$41:36:59.4 & 0.10 \\
Cyg OB2-75 & O9V & (2) & 20:34:09.52,$+$41:34:13.7 & 0.10 \\
Cyg OB2-29 & O7.5V(n)((f))z & (2) & 20:34:13.51,$+$41:35:03.0 & 0.10 \\
J20341605$+$4102196 & O9.5V & (4) & 20:34:16.05,$+$41:02:19.6 & 0.11 \\
J20342894$+$4156171 & O9V & (4) & 20:34:28.94,$+$41:56:17.0 & 0.10 \\
Cyg OB2-A24 & O6.5III(f) & (2) & 20:34:44.11,$+$40:51:58.5 & 0.16 \\
CPR2002-A10 & O9V & (9) & 20:34:55.11,$+$40:34:44.3 & 0.30 \\
ALS 19627 & O9.7Iab & (4) & 20:34:56.06,$+$40:38:17.9 & 0.15 \\
2MASSI\,J2035238$+$412203 & O8.5V & (10) & 20:35:23.79,$+$41:22:03.7 & 0.15  \\
ALS 19631 & O5V((f)) & (2) & 20:36:04.50,$+$40:56:13.0 & 0.25\\
\hline \hline
\end{tabular*}
\tabnote{References: (1): \citet{Mahy2013}; 
(2): Galactic O-star catalog v. 4.2 (http://gosc.cab.inta-csic.es/gosc.php); 
(3): \citet{gvaramadze2014};
(4): \citet{comeron2012}; 
(5): \citet{hoag1965};
(6): \citet{sota2011};
(7): \citet{kiminki2007};
(8): \citet{simondiaz2007}; 
(9): \citet{kobulnicky2010};
(10): \citet{2MASScat}.}
\end{table*}

\end{appendix}

\end{document}